\documentclass[lettersize,journal]{IEEEtran}

\usepackage{amsmath,amsfonts}
\usepackage{algorithm}
\usepackage{array}
\usepackage{textcomp}
\usepackage{stfloats}
\usepackage{url}
\usepackage{verbatim}
\usepackage{graphicx}
\usepackage{cite}

\usepackage{xcolor}
\usepackage{amssymb}
\usepackage{algpseudocode}
\usepackage{subcaption}
\usepackage{comment}

\begin{document}
\title{Radio Map Construction with Post-Hoc Location Calibration under Quasi-Static Positioning Errors: Joint Estimation, Performance Bounds, and GNSS-Based Evaluation}

\author{Koki Kanzaki,~\IEEEmembership{Graduate Student Member, IEEE}, Katsuya Suto,~\IEEEmembership{Senior Member, IEEE} \\ and Koya Sato,~\IEEEmembership{Senior Member, IEEE}
\thanks{A part of this work was presented at the IEEE GLOBECOM Workshops\,\cite{kanzakiJointExPostLocation2025} (DOI: 10.1109/GCWkshps68340.2025.11591047).}
\thanks{This work was supported by JST, PRESTO under Grant Number JPMJPR23P3, CRONOS under Grant Number JPMJCS24N1, ASPIRE under Grant JPMJAP2346, BOOST under Grant Number JPMJBS2415, and JSPS KAKENHI under Grant Number 25K00138.}%
\thanks{Corresponding author: Koki Kanzaki.}
\thanks{Koki Kanzaki and Koya Sato are with the Artificial Intelligence eXploration Research Center, The University of Electro-Communications, Tokyo 182-8585, Japan (e-mail: k-kanzaki@uec.ac.jp and k\_sato@ieee.org).}
\thanks{Katsuya Suto is with the Faculty of Information Science and Technology, Hokkaido University, Sapporo, Hokkaido 060-0808, Japan (e-mail: k.suto@ist.hokudai.ac.jp).}
}

\markboth{Submitted to IEEE Internet of Things Journal, September~2026}%
{Shell \MakeLowercase{\textit{et al.}}: A Sample Article Using IEEEtran.cls for IEEE Journals}

\maketitle

\begin{abstract}
Radio maps enable environment-aware wireless and Internet-of-Things applications and can be constructed from location-tagged received signal strength (RSS) measurements collected by mobile devices. In urban environments, temporally correlated GNSS errors can shift an entire sensing trajectory, causing systematic spatial misregistration that is not mitigated by collecting more measurements.
This paper presents a radio-map construction framework that uses the radio measurements themselves to calibrate erroneous location tags after data collection. The dominant positioning error is modeled as a sensor-specific quasi-static offset, which is jointly estimated with radio-propagation parameters in a Gaussian process regression (GPR) framework by exploiting complementary spatial information from distance-dependent path loss and spatially correlated shadowing. We establish lower and upper bounds on the conditional Bayes risk and show that, under a translation-invariant trajectory model, trajectory information alone cannot identify the quasi-static offset, thereby motivating the use of RSS-derived spatial information for calibration. Numerical evaluations across propagation conditions show that the proposed method reduces the mean squared error (MSE) gap from ideal GPR to approximately $3.26\,\mathrm{dB}^2$, compared with about $10\,\mathrm{dB}^2$ for position-error-agnostic and noisy-input GPR baselines. Evaluation using positioning-error models derived from smartphone GNSS measurements shows that the proposed method outperforms a KF--RTS trajectory-smoothing baseline despite unmodeled time-varying positioning errors, remaining within approximately $5\,\mathrm{dB}^2$ of ideal GPR at the median MSE. These results demonstrate that RSS measurements can serve not only as observations for radio-map reconstruction but also as spatial cues for post-hoc calibration of imperfectly geotagged sensing data.
\end{abstract}


\section{Introduction}
\label{sec:introduction}
\IEEEPARstart{E}{nvironment}-aware wireless communication requires site-specific knowledge of radio-propagation conditions.
Radio maps provide such spatial knowledge by associating radio-propagation quantities, such as received signal strength (RSS), path loss, and channel gain, with geographical locations\,\cite{romeroRadioMapEstimation2022b}.
They have been widely investigated as a basis for wireless resource management, localization, network optimization, and other environment-aware applications\,\cite{biEngineeringRadioMaps2019b,zhangWiFiBasedIndoorLocalization2024a,suarezrodriguezNetworkOptimisation5G2020}.
More broadly, spatial representations of radio-environment knowledge are also important components of network digital twins and channel knowledge maps (CKMs) for next-generation wireless systems\,\cite{tranNetworkDigitalTwin2025,wangDigitalTwinChannel2025,zengTutorialEnvironmentAwareCommunications2024b}.
\par
Measurement-based radio-map construction is particularly attractive because it can directly capture site-specific propagation characteristics.
In a typical mobile-sensing architecture, terminals such as smartphones collect radio measurements while moving, associate each measurement with an estimated location, and upload the resulting data to a remote server.
The server aggregates measurements collected by multiple terminals and estimates the radio conditions at unobserved locations through spatial interpolation.
A variety of model-driven and data-driven estimators have been developed for this purpose\,\cite{chenGPRTGaussianProcess2025,sunPropagationMapReconstruction2022,teganyaDeepCompletionAutoencoders2022}.
Among them, Gaussian process regression (GPR)\,\cite{rasmussenGaussianProcessesMachine2008} provides a probabilistic interpolation framework that can explicitly exploit the spatial correlation of shadowing to reconstruct radio maps from sparse measurements.
\par
A fundamental difficulty in such measurement-based mapping is that the measurement coordinates themselves can be inaccurate.
This is particularly relevant to outdoor mobile sensing, where measurement locations are commonly obtained using a Global Navigation Satellite System (GNSS).
In urban environments, surrounding buildings, non-line-of-sight (NLOS) reception, multipath propagation, and satellite geometry can substantially degrade GNSS positioning accuracy\,\cite{zhuGNSSPositionIntegrity2018}.
Moreover, GNSS positioning errors are not necessarily independent across time.
When the propagation conditions of GNSS signals or the geometry of the received satellites remain similar over a certain period, the resulting positioning displacement can persist across consecutive measurements.
Indeed, positioning errors observed along mobile trajectories have been reported to exhibit strong temporal autocorrelation\,\cite{xiongFaultTolerantGNSSSINS2020a,ranacherWhyGPSMakes2016}.
Consequently, a sequence of measurements collected by the same terminal can be coherently displaced from the true trajectory rather than independently scattered around it.
\par
In this paper, we focus on the dominant quasi-static component of such temporally correlated positioning errors, which we refer to as a \emph{quasi-static positioning bias}.
Specifically, we model this component as a position offset shared by the measurements along each trajectory.
In the mobile-sensing setting considered in this paper, each sensor generates one measurement trajectory and is therefore associated with one shared positioning offset.
\par
\begin{figure}[t!]
    \centering
    \subfloat[]{
        \includegraphics[width=0.45\linewidth]{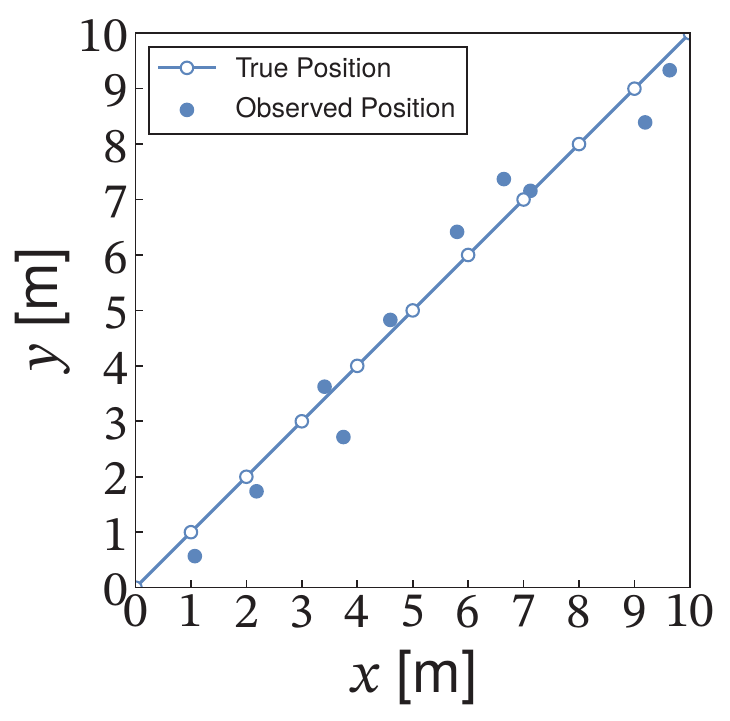}
        \label{fig:iid_positioning_error}
    }
    \hfill
    \subfloat[]{
        \includegraphics[width=0.45\linewidth]{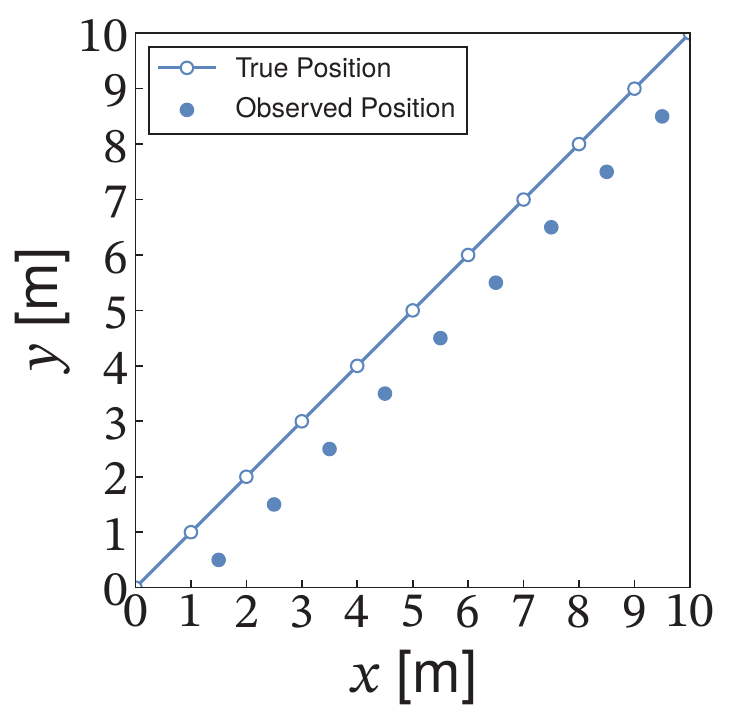}
        \label{fig:burst_positioning_error}
    }
    \caption{Illustrative examples of positioning errors:
    (a) i.i.d. errors and
    (b) trajectory-wise quasi-static bias.
    }
    \label{fig:examples_positioning_errors}
\end{figure}
Fig.\,\ref{fig:examples_positioning_errors} highlights an important distinction between pointwise positioning uncertainty and the trajectory-wise bias.
In Fig.\,\ref{fig:examples_positioning_errors}(a), independent positioning errors perturb individual measurements around the true trajectory.
In contrast, in Fig.\,\ref{fig:examples_positioning_errors}(b), a sequence of reported positions is coherently displaced in a common direction.
Such a shared offset is not averaged out by collecting additional measurements along the same trajectory.
Instead, multiple radio measurements remain systematically registered at incorrect spatial locations, which can directly produce spatial misregistration in the reconstructed radio map.
\par
Two classes of existing approaches are particularly relevant to this problem.
The first accounts for uncertainty in measurement locations within the regression model.
For example, noisy-input Gaussian processes (NIGPs) approximate the effect of Gaussian input noise as additional output uncertainty\,\cite{mchutchonGaussianProcessTraining2011a}, while more general GP-based methods explicitly incorporate uncertain or latent inputs into the inference procedure\,\cite{girardGaussianProcessPriors2002,damianouVariationalInferenceLatent2016}.
For radio-map construction, positional uncertainty has also been incorporated into GP inference through variational methods\,\cite{zhenRadioEnvironmentMap2022}.
These approaches provide principled mechanisms for handling uncertainty in individual input locations.
However, the trajectory-wise quasi-static bias is qualitatively different: a common offset coherently translates a sequence of measurements along the same trajectory.
Thus, treating such a shared displacement only as pointwise input uncertainty does not remove the resulting trajectory-level spatial misregistration.
\par
The second class of approaches attempts to improve the positioning trajectory before radio-map construction.
Trajectory-estimation methods based on the Kalman filter (KF), the Rauch--Tung--Striebel (RTS) smoother, and GNSS/IMU fusion can exploit motion models and temporal error correlation to suppress time-varying positioning errors\,\cite{kalmanNewApproachLinear1960,sarkkaBayesianFilteringSmoothing2023,iyerEnhancingPositioningGNSS2024}.
However, trajectory information alone does not necessarily determine a trajectory-wise constant positioning offset.
Without sufficiently informative absolute-location constraints, a constant translation of the underlying trajectory can be compensated for by an opposite change in the positioning bias while producing the same reported trajectory.
Thus, temporal smoothing can reduce time-varying positioning errors while leaving a residual absolute spatial misregistration that directly affects subsequent radio-map construction.
\par
The key observation of this work is that the radio measurements themselves provide spatial information that is unavailable to trajectory-only correction.
The mean RSS depends on the transmitter--receiver distance through distance-dependent path loss and therefore provides information about the absolute placement of a measurement trajectory relative to a known base station (BS).
At the same time, spatially correlated shadowing provides pairwise information about the relative placement of measurements collected by different sensors.
Hence, RSS observations can provide additional spatial constraints for estimating trajectory-wise positioning offsets.
From this perspective, radio-map construction and location calibration are coupled inference problems: the measurements used to reconstruct the radio environment can also help determine where those measurements were actually collected.
\par
Motivated by this observation, we propose a GPR-based radio-map construction framework with post-hoc location calibration under quasi-static positioning biases.
The proposed method models the dominant positioning error as an unknown offset shared by the measurements along each trajectory and jointly estimates these offsets and the radio-propagation parameters from the RSS observations.
The estimation exploits both the distance-dependent mean and the spatial covariance structure of RSS.
The estimated offsets are then used to calibrate the measurement coordinates, and the calibrated locations are used as the GPR inputs for radio-map construction.
In this way, the same RSS measurements serve both as observations for reconstructing the radio map and as spatial information for post-hoc calibration of their associated location tags.
\par
The main contributions of this work are summarized as follows.
\begin{itemize}
    \item
    We develop a GPR-based radio-map construction framework that jointly performs post-hoc location calibration and radio-map estimation under trajectory-wise quasi-static positioning biases.
    By exploiting distance-dependent path loss and spatially correlated shadowing, the proposed method jointly estimates the positioning offsets and radio-propagation parameters and reconstructs the radio map using the calibrated measurement locations.

    \item
    We characterize the fundamental role of radio observations in resolving trajectory-wise positioning offsets.
    We show that, under a translation-invariant trajectory law, the reported positioning trajectory alone does not update the prior distribution of a trajectory-wise constant offset, clarifying why additional spatial information is required for post-hoc calibration.
    We further establish a bounding framework based on a conditional minimum-mean-square-error (MMSE) benchmark, deriving lower and upper bounds on the conditional Bayes risk and thereby bounding the excess risk of an arbitrary radio-map estimator without explicitly evaluating the Bayes-optimal estimator.

    \item
    We evaluate the proposed method over a broad range of radio-propagation conditions and further assess its robustness to positioning-error model mismatch.
    Using positioning-error models derived from measured smartphone GNSS data, we consider temporally correlated error components that are not explicitly represented by the proposed quasi-static offset model.
    In addition to position-error-agnostic and uncertain-input methods, we compare the proposed method with a KF--RTS-based trajectory-correction baseline that explicitly exploits temporal error correlation, thereby assessing the benefit of using the spatial structure of the radio observations themselves for location calibration.
\end{itemize}
Sec.\,\ref{sec:system_model} describes the system model.
Secs.\,\ref{sec:without_accounting} and
\ref{sec:position_error_aware} present the baseline and proposed
radio-map construction methods, respectively.
Sec.\,\ref{sec:performance_bounds} presents the bounding framework
for evaluating the deviation from Bayes-optimal performance.
Sec.\,\ref{sec:numerical_evaluation} evaluates the proposed method
under various radio-propagation conditions.
Sec.\,\ref{sec:real_gnss_evaluation} further evaluates its robustness
to temporally correlated positioning-error components
that are not explicitly represented by the quasi-static offset model,
using error models derived from measured GNSS data.
Finally, Sec.\,\ref{sec:conclusion} concludes this paper. \par
{\it Notations:}
The transpose and inverse are denoted by $(\cdot)^\top$ and $(\cdot)^{-1}$, respectively, and $\lvert \mathbf{A} \rvert$ denotes the determinant of a square matrix $\mathbf{A}$.
The Euclidean norm is denoted by $\lVert \cdot \rVert$, and $\operatorname{tr}(\cdot)$ denotes the trace.
$\mathbb{E}[\cdot]$ and $\operatorname{Cov}[\cdot]$ denote expectation and covariance, respectively.
In Sec.\,\ref{sec:performance_bounds}, $\mathbb{E}_{\boldsymbol{\theta}}[\cdot]$ and $\operatorname{Cov}_{\boldsymbol{\theta}}(\cdot)$ denote expectation and covariance under the generative model with $\boldsymbol{\theta}$ fixed at its true value.

\section{System Model}
\label{sec:system_model}
\begin{figure}[t!]
    \centering
    \includegraphics[width=\linewidth]{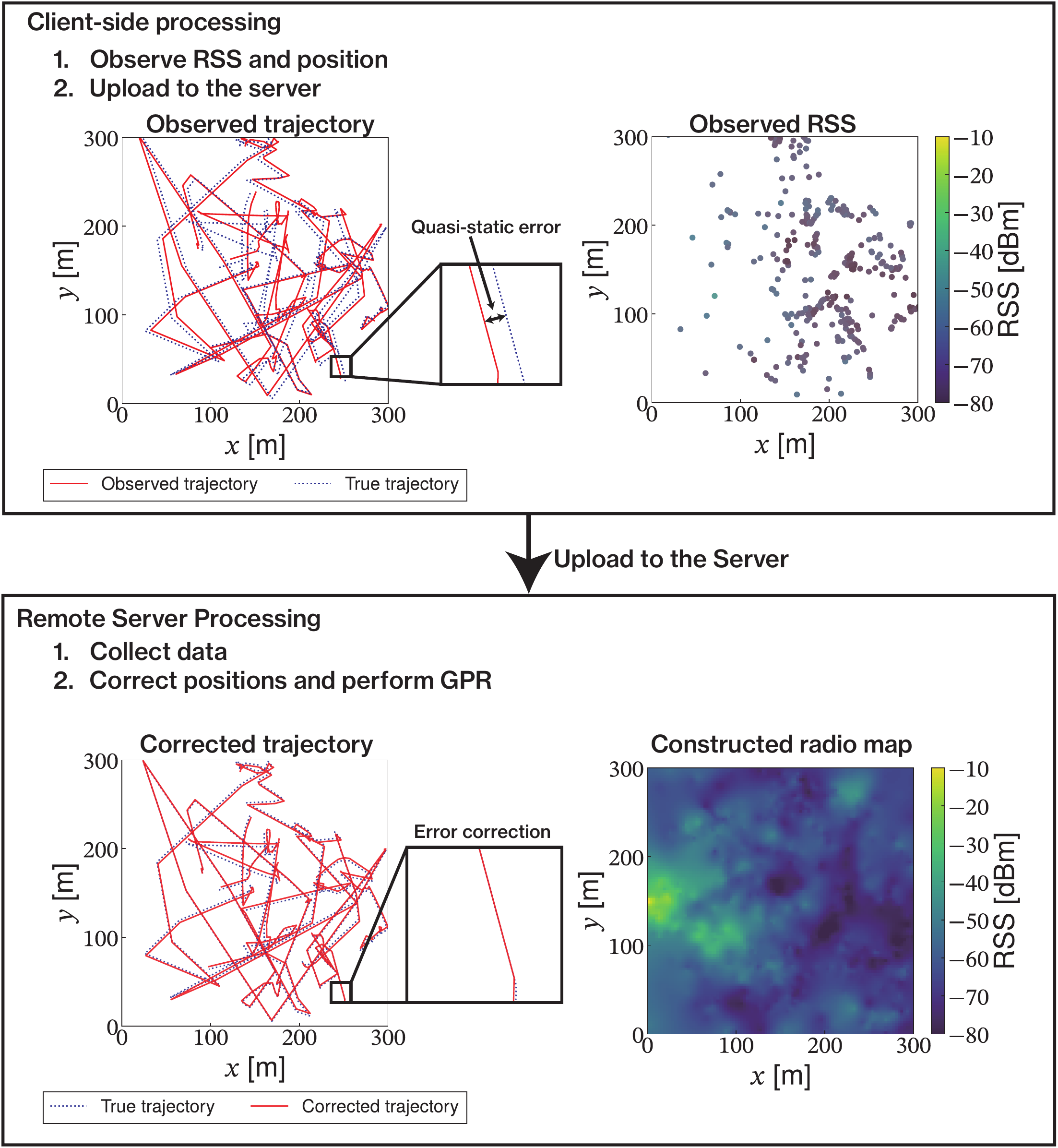}
    \caption{Overview of the system model.}
    \label{fig:overview_of_system_model}
\end{figure}
Fig.\,\ref{fig:overview_of_system_model} provides an overview of the system model.
We consider the construction of a radio map that estimates the received power from a single base station (BS) at each location.
In the target area, $N$ mobile devices measure the received power while moving and report the measurements to a remote server together with their corresponding observed coordinates.
After collecting sufficient measurements, the server constructs a radio map by spatially interpolating the RSS at unobserved locations.
The radio propagation parameters are also estimated during this process. \par
Let the target area be a two-dimensional region $\mathcal{A} \subset \mathbb{R}^2$, and let $\mathbf{x}_{\mathrm{Tx}} \in \mathcal{A}$ denote the coordinates of the BS, which are assumed to be known.
Assuming that the effects of small-scale fading are largely suppressed through signal averaging during the measurement process, the received power $P_{\mathrm{Rx}}\,[\mathrm{dBm}]$ at $\mathbf{x} \in \mathcal{A}$ can be modeled as\,\cite{goldsmithWirelessCommunications2005},
\begin{align}
    \label{eq:received_power_model}
    P_{\mathrm{Rx}}(\mathbf{x}) &= P_{\mathrm{Tx}} - 10 \eta \log_{10}\left(\frac{\lVert \mathbf{x}_{\mathrm{Tx}} - \mathbf{x} \rVert}{d_0}\right) + W(\mathbf{x}) \\
    &\triangleq f(\mathbf{x}),
\end{align}
where $P_{\mathrm{Tx}}\,[\mathrm{dBm}]$ denotes the transmit power of the BS, $\eta$ denotes the path-loss index, $\lVert\cdot\rVert$ denotes the Euclidean distance, and $W(\cdot)$ denotes the shadowing term.
The constant $d_0$ is a reference distance.
Following Gudmundson's shadowing correlation model\,\cite{gudmundsonCorrelationModelShadow1991}, the shadowing component can be modeled as
\begin{equation}
    W(\cdot) \sim \mathrm{GP}\left(0, k_{\mathrm{exp}}(\cdot, \cdot)\right),
\end{equation}
where $k_{\mathrm{exp}}(\cdot, \cdot)$ is the exponential kernel defined as
\begin{equation}
    \label{eq:exp_kernel}
    k_{\mathrm{exp}}\left( \mathbf{x}, \mathbf{x}' \right) = \sigma_f^2 \exp \left( - \frac{\lVert \mathbf{x} - \mathbf{x}' \rVert \ln{2}}{d_{\mathrm{cor}}} \right).
\end{equation}
Here, $\sigma_f^2$ is the variance of the shadowing term, and $d_{\mathrm{cor}}$ is the correlation distance. \par
Let $M^{(i)}$ denote the number of measurements collected by the $i$-th sensor, and let $\mathbf{x}^{(i)}_j$ denote its location at time index $j$, for $j = 1, \cdots, M^{(i)}$.
The corresponding RSS measurement, denoted by $p_j^{(i)}$, is modeled as
\begin{equation}
    \label{eq:rss_observation}
    p^{(i)}_j = f\left( \mathbf{x}^{(i)}_j \right) + \epsilon^{(i)}_{p, j}, \quad \epsilon^{(i)}_{p, j} \sim \mathcal{N}\left(0, \sigma_p^2\right),
\end{equation}
where $\epsilon^{(i)}_{p,j}$ denotes Gaussian measurement noise that captures thermal noise and residual small-scale fading remaining after signal averaging and $\sigma_p^2$ is its variance. \par
We next model the errors in the reported coordinates.
As discussed in Sec.\,\ref{sec:introduction}, GNSS positioning errors can exhibit strong temporal persistence.
As a tractable approximation, we model their dominant quasi-static component as a sensor-specific constant offset over the observation interval.
Residual time-varying positioning errors that are not captured by this model are considered in Sec.\,\ref{sec:real_gnss_evaluation}.
Let $\tilde{\mathbf{x}}_j^{(i)}$ denote the position reported by sensor $i$ at time index $j$.
Using the true position $\mathbf{x}_j^{(i)}$, the observed position is modeled as
\begin{equation}
    \label{eq:noisy_observation}
    \tilde{\mathbf{x}}_j^{(i)} = \mathbf{x}_j^{(i)} + \mathbf{e}^{(i)}.
\end{equation}
Under this approximation, $\mathbf{e}^{(i)}$ is assumed to remain constant over the observation interval of the $i$-th sensor and to be independent across sensors.
Further, it follows
\begin{equation}
    \label{eq:position_error_prior}
    \mathbf{e}^{(i)} \sim \mathcal{N}(\boldsymbol{\mu}_s, \boldsymbol{\Sigma}_s),
\end{equation}
where $\boldsymbol{\mu}_s \in \mathbb{R}^2$ and $\boldsymbol{\Sigma}_s \in \mathbb{R}^{2 \times 2}$ denote the mean vector and the covariance matrix of the positioning error, respectively\footnote{The prior covariance can be obtained from positioning-quality indicators or estimated empirically from positioning-error statistics.}.
We further assume that $\{\mathbf{e}^{(i)}\}_{i=1}^{N}$ is independent of the shadowing field, of the measurement noise, and of the true measurement locations. \par
Based on the above observation model, we define the measurement data $\mathcal{D}^{(i)}$ collected by the $i$-th sensor as
\begin{equation}
    \mathcal{D}^{(i)} \triangleq \left\{ \left( \tilde{\mathbf{x}}_j^{(i)}, p^{(i)}_j \right)\,\middle|\, j = 1, \cdots, M^{(i)} \right\}.
\end{equation}
After all sensors upload their local dataset, the server aggregates the local datasets as
\begin{equation}
    \mathcal{D} \triangleq \left\{ \mathcal{D}^{(i)} \right\}_{i=1}^N.
\end{equation}
The server is assumed to know the transmitter coordinates $\mathbf{x}_{\mathrm{Tx}}$, the transmit power $P_{\mathrm{Tx}}$, and the prior statistics $\boldsymbol{\mu}_s$ and $\boldsymbol{\Sigma}_s$ of the positioning errors.
Let $\mathcal{X}_\ast \subseteq \mathcal{A}$ denote the set of query points at which the radio map is evaluated, and let $\mathbf{x}_\ast \in \mathcal{X}_\ast$ denote a generic query point.
Given these quantities and the dataset $\mathcal{D}$, the server estimates the RSS at $\mathbf{x}_\ast$.

\section{Radio Map Construction without Accounting for Position Errors}
\label{sec:without_accounting}
We first establish a position-error-agnostic GPR baseline that treats the reported measurement locations as the true GP inputs.
This approach treats the reported measurement locations as the true locations; i.e., $\mathbf{x}_j^{(i)} \approx \tilde{\mathbf{x}}_{j}^{(i)}$.
\par
Ignoring positioning error, the unknown model parameters in the system model are collected into
\begin{equation}
    \label{eq:ordinary_parameter_vector}
    \boldsymbol{\theta} \triangleq \left[ \eta, \sigma_f^2, d_{\mathrm{cor}}, \sigma_p^2 \right]^\top.
\end{equation}
Because the shadowing component is modeled as a zero-mean GP, the latent field of RSS can be expressed as
\begin{equation}
    f(\mathbf{x}) \sim \mathrm{GP}\left( m_{\boldsymbol{\theta}}(\mathbf{x}), k_{\mathrm{exp}, \boldsymbol{\theta}}(\mathbf{x}, \mathbf{x}') \right),
\end{equation}
where the mean function is given by
\begin{equation}
    \label{eq:ordinary_mean_function}
    m_{\boldsymbol{\theta}}(\mathbf{x}) \triangleq P_{\mathrm{Tx}} - 10\eta \log_{10}\left(1 + \frac{\left\lVert \mathbf{x}_{\mathrm{Tx}} - \mathbf{x} \right\rVert}{d_0}\right),
\end{equation}
and the exponential kernel is given by Eq.\,\eqref{eq:exp_kernel} with the corresponding parameters specified by $\boldsymbol{\theta}$\footnote{The additive constant $1$ is introduced for analytical convenience. Its effect is negligible over the communication distance considered in this paper, for which $\lVert \mathbf{x}_{\mathrm{Tx}} - \mathbf{x}\rVert / d_0 \gg 1$.}.\par
Let us define the total number of measurements collected from all sensors as
\begin{equation}
    M_N \triangleq \sum_{i = 1}^N M^{(i)}.
\end{equation}
To express the GP likelihood in a vector form, we enumerate all measurements using a global index $n=1, \cdots, M_N$.
For the $n$-th measurement, let $i_n$ denote the sensor index and $j_n$ denote the time index within the measurement sequence of sensor $i_n$.
Then, the reported location $\tilde{\mathbf{x}}_n$ and the RSS measurement $p_n$ are defined as
\begin{align}
    \tilde{\mathbf{x}}_n &\triangleq \tilde{\mathbf{x}}_{j_n}^{(i_n)}, \\
    p_n &\triangleq p^{(i_n)}_{j_n}.
\end{align}
The reported locations and RSS measurements are vectorized as
\begin{align}
    \tilde{\mathbf{X}} &\triangleq \left[ \tilde{\mathbf{x}}_1, \cdots, \tilde{\mathbf{x}}_{M_N} \right]^\top
    \in \mathbb{R}^{M_N \times 2}, \\
    \mathbf{p} &\triangleq \left[p_1, \cdots, p_{M_N}\right]^\top \in \mathbb{R}^{M_N},
\end{align}
respectively.
The observation vector conditioned on the reported locations and model parameters follows
\begin{equation}
    \mathbf{p} \mid \tilde{\mathbf{X}}, \boldsymbol{\theta} \sim \mathcal{N}\left( \mathbf{m}_{\boldsymbol{\theta}, \tilde{\mathbf{X}}}, \mathbf{C}_{\boldsymbol{\theta}, \tilde{\mathbf{X}}} \right),
\end{equation}
where the mean vector is
\begin{equation}
    \label{eq:ordinary_mean_vector}
    \mathbf{m}_{\boldsymbol{\theta}, \tilde{\mathbf{X}}} \triangleq \left[ m_{\boldsymbol{\theta}}(\tilde{\mathbf{x}}_1), \cdots, m_{\boldsymbol{\theta}}(\tilde{\mathbf{x}}_{M_N}) \right]^\top,
\end{equation}
and the covariance matrix is
\begin{equation}
    \label{eq:ordinary_covariance_matrix}
    \mathbf{C}_{\boldsymbol{\theta}, \tilde{\mathbf{X}}} \triangleq \mathbf{K}_{\boldsymbol{
    \theta}, \tilde{\mathbf{X}}} + \sigma_p^2 \mathbf{I}.
\end{equation}
Here, the kernel matrix $\mathbf{K}_{\boldsymbol{\theta}, \tilde{\mathbf{X}}}$ has entries
\begin{equation}
    \label{eq:ordinary_kernel_matrix}
    \left[ \mathbf{K}_{\boldsymbol{\theta}, \tilde{\mathbf{X}}}\right]_{m,n} = k_{\mathrm{exp}, \boldsymbol{\theta}} \left( \tilde{\mathbf{x}}_m, \tilde{\mathbf{x}}_n \right), \quad m,n = 1, \cdots, M_N.
\end{equation}
The resulting log marginal likelihood is
\begin{multline}
    \label{eq:ordinary_marginal_log_likelihood}
    \log \mathcal{L} \left( \mathbf{p} \mid \tilde{\mathbf{X}}, \boldsymbol{\theta} \right) = -\frac{1}{2} \log \left\lvert \mathbf{C}_{\boldsymbol{\theta}, \tilde{\mathbf{X}}} \right\rvert - \frac{M_N}{2} \log(2\pi) \\
    -\frac{1}{2} \left( \mathbf{p} - \mathbf{m}_{\boldsymbol{\theta}, \tilde{\mathbf{X}}} \right)^\top \mathbf{C}_{\boldsymbol{\theta}, \tilde{\mathbf{X}}}^{-1} \left( \mathbf{p} - \mathbf{m}_{\boldsymbol{\theta}, \tilde{\mathbf{X}}} \right).
\end{multline}
The model parameters are estimated by maximizing the log marginal likelihood:
\begin{equation}
    \label{eq:ordinary_parameter_estimation}
    \hat{\boldsymbol{\theta}} = \underset{\boldsymbol{\theta} \in \Theta}{\operatorname{arg\,max}}\; \log \mathcal{L}\left( \mathbf{p} \mid \tilde{\mathbf{X}}, \boldsymbol{\theta} \right),
\end{equation}
where $\Theta$ denotes the feasible parameter set.
We can maximize this function based on a gradient-based method; this paper solves this problem using Adam\,\cite{kingmaAdamMethodStochastic2017}. \par
Using the estimated parameters, the cross-covariance vector between the query point $\mathbf{x}_\ast$ and the reported measurement locations is given by
\begin{equation}
    \label{eq:ordinary_cross_covariance_vector}
    \left[ \mathbf{k}_{\hat{\boldsymbol{\theta}}, \tilde{\mathbf{X}}}(\mathbf{x}_\ast) \right]_n = k_{\mathrm{exp}, \hat{\boldsymbol{\theta}}} \left( \mathbf{x}_\ast, \tilde{\mathbf{x}}_n \right), \quad n = 1, \cdots, M_N
\end{equation}
The RSS at $\mathbf{x}_\ast$ is then estimated by the posterior mean of the latent field of RSS:
\begin{align}
    \hat{p}(\mathbf{x}_\ast) &\triangleq \mathbb{E}\left[ f(\mathbf{x}_\ast) \mid \mathbf{p}, \tilde{\mathbf{X}}, \hat{\boldsymbol{\theta}} \right] \\
    \label{eq:ordinary_gp_posterior_mean}
    &= m_{\hat{\boldsymbol{\theta}}}(\mathbf{x}_\ast) + \mathbf{k}_{\hat{\boldsymbol{\theta}}, \tilde{\mathbf{X}}}^\top(\mathbf{x}_\ast) \mathbf{C}_{\hat{\boldsymbol{\theta}}, \tilde{\mathbf{X}}}^{-1} \left( \mathbf{p} - \mathbf{m}_{\hat{\boldsymbol{\theta}}, \tilde{\mathbf{X}}} \right).
\end{align}
Evaluating this posterior mean over the query points yields the radio map.
The complete procedure is summarized in Alg.\,\ref{alg:position_error_agnostic}.

\begin{algorithm}[t]
  \caption{Radio map construction without accounting for positioning errors}
  \label{alg:position_error_agnostic}
  \begin{algorithmic}[1]
    \State \textbf{Input:}
    Dataset $\mathcal{D}$,
    transmitter location $\mathbf{x}_{\mathrm{Tx}}$,
    transmit power $P_{\mathrm{Tx}}$,
    and query-point set $\mathcal{X}_{\ast}$

    \State \textbf{Output:}
    Radio map $\{\hat{p}(\mathbf{x}_{\ast})\}_{\mathbf{x}_{\ast}\in\mathcal{X}_{\ast}}$ and estimated model parameters $\hat{\boldsymbol{\theta}}$
    \State Form the reported-location matrix $\tilde{\mathbf{X}}$ and RSS observation vector $\mathbf{p}$ from $\mathcal{D}$
    \State Estimate $\hat{\boldsymbol{\theta}}$ by solving Eq.\,\eqref{eq:ordinary_parameter_estimation}
    \State Form $\mathbf{m}_{\hat{\boldsymbol{\theta}}, \tilde{\mathbf{X}}}$
    using Eq.\,\eqref{eq:ordinary_mean_vector}
    \State Form $\mathbf{K}_{\hat{\boldsymbol{\theta}}, \tilde{\mathbf{X}}}$ using Eq.\,\eqref{eq:ordinary_kernel_matrix} and $\mathbf{C}_{\hat{\boldsymbol{\theta}}, \tilde{\mathbf{X}}}$ using Eq.\,\eqref{eq:ordinary_covariance_matrix}
    \ForAll{$\mathbf{x}_{\ast}\in\mathcal{X}_{\ast}$}
      \State Form $\mathbf{k}_{\hat{\boldsymbol{\theta}}, \tilde{\mathbf{X}}}(\mathbf{x}_\ast)$ using Eq.\,\eqref{eq:ordinary_cross_covariance_vector}
      \State Compute $\hat{p}(\mathbf{x}_\ast)$ using Eq.\,\eqref{eq:ordinary_gp_posterior_mean}
    \EndFor
    \State \Return $\{\hat{p}(\mathbf{x}_{\ast})\}_{\mathbf{x}_{\ast}\in\mathcal{X}_{\ast}}$ and $\hat{\boldsymbol{\theta}}$
  \end{algorithmic}
\end{algorithm}

\section{Radio Map Construction and Joint Estimation of Propagation Parameters and Position Errors}
\label{sec:position_error_aware}
Alg.\,\ref{alg:position_error_agnostic} treats the reported measurement locations as exact GP inputs.
We now relax this assumption by treating the sensor-specific positioning offsets as unknown variables jointly estimated with the propagation parameters.
Because these offsets shift the GP inputs, they affect both the distance-dependent mean and the spatial covariance of the RSS observations.
The proposed method exploits these two spatial structures to jointly estimate the model parameters and the sensor-specific positioning errors.

\subsection{Joint Estimation of Propagation Parameters and Position Errors}
From the positioning error model in Eq.\,\eqref{eq:noisy_observation}, the true location of the $j$-th measurement collected by sensor $i$ is given by
\begin{equation}
    \label{eq:noise_model}
    \mathbf{x}_j^{(i)} = \tilde{\mathbf{x}}_j^{(i)} - \mathbf{e}^{(i)}.
\end{equation}
We collect the sensor-specific positioning errors into the vector
\begin{equation}
    \label{eq:stacked_position_error}
    \mathbf{e}_{\mathrm{vec}} \triangleq \left[ \left(\mathbf{e}^{(1)}\right)^\top, \cdots, \left(\mathbf{e}^{(N)}\right)^\top \right]^\top \in \mathbb{R}^{2N}.
\end{equation}
Under the global measurement index introduced in the previous section, let
\begin{equation}
    \mathbf{e}_n \triangleq \mathbf{e}^{(i_n)} \quad n = 1, \cdots, M_N.
\end{equation}
The corrected location associated with the $n$-th measurement is then
\begin{equation}
    \label{eq:corrected_stacked_location}
    \mathbf{x}_n(\mathbf{e}_{\mathrm{vec}}) \triangleq \tilde{\mathbf{x}}_n - \mathbf{e}_n.
\end{equation}
Collecting these locations gives
\begin{equation}
    \label{eq:corrected_location_matrix}
    \mathbf{X}(\mathbf{e}_{\mathrm{vec}}) \triangleq \left[ \mathbf{x}_1(\mathbf{e}_{\mathrm{vec}}), \cdots, \mathbf{x}_{M_N}(\mathbf{e}_{\mathrm{vec}}) \right]^\top
\end{equation}
Conditioned on the reported locations, model parameters, and positioning errors, the RSS observation vector is Gaussian with mean
\begin{multline}
    \label{eq:position_error_aware_mean_vector}
    \mathbf{m}_{\boldsymbol{\theta},\mathbf{X}(\mathbf{e}_{\mathrm{vec}})} \triangleq \left[ m_{\boldsymbol{\theta}} \left(\mathbf{x}_1(\mathbf{e}_{\mathrm{vec}})\right), \cdots, m_{\boldsymbol{\theta}} \left(\mathbf{x}_{M_N}(\mathbf{e}_{\mathrm{vec}})\right) \right]^\top
\end{multline}
Its $n$-th element can be written explicitly as
\begin{equation}
    \left[ \mathbf{m}_{\boldsymbol{\theta}, \mathbf{X}(\mathbf{e}_{\mathrm{vec}})} \right]_n = P_{\mathrm{Tx}} - 10\eta \log_{10} \left( 1 + \frac{\left\lVert \mathbf{x}_{\mathrm{Tx}} - \mathbf{x}_n(\mathbf{e}_{\mathrm{vec}}) \right\rVert}{d_0} \right).
\end{equation}
The corresponding covariance matrix is
\begin{equation}
    \label{eq:position_error_aware_covariance_matrix}
    \mathbf{C}_{\boldsymbol{\theta}, \mathbf{X}(\mathbf{e}_{\mathrm{vec}})} \triangleq \mathbf{K}_{\boldsymbol{\theta}, \mathbf{X}(\mathbf{e}_{\mathrm{vec}})} + \sigma_p^2\mathbf{I}
\end{equation}
where the kernel matrix $\mathbf{K}_{\boldsymbol{\theta}, \mathbf{X}(\mathbf{e}_{\mathrm{vec}})}$ has entries
\begin{align}
    &\left[ \mathbf{K}_{\boldsymbol{\theta}, \mathbf{X}(\mathbf{e}_{\mathrm{vec}})} \right]_{m,n} \notag\\
    &\quad = k_{\mathrm{exp},\boldsymbol{\theta}} \left( \mathbf{x}_m(\mathbf{e}_{\mathrm{vec}}), \mathbf{x}_n(\mathbf{e}_{\mathrm{vec}}) \right) \notag\\
    \label{eq:position_error_aware_kernel_matrix}
    &\quad = \sigma_f^2 \exp \left( - \frac{ \left\lVert (\tilde{\mathbf{x}}_m-\mathbf{e}_m) - (\tilde{\mathbf{x}}_n-\mathbf{e}_n) \right\rVert \ln(2)}{d_{\mathrm{cor}}} \right),
\end{align}
for $m,n = 1, \cdots, M_N$.
The resulting log marginal likelihood is
\begin{multline}
    \label{eq:MLE}
    \log \mathcal{L} \left( \mathbf{p} \mid \tilde{\mathbf{X}}, \boldsymbol{\theta}, \mathbf{e}_{\mathrm{vec}} \right) = - \frac{1}{2} \log \left\lvert \mathbf{C}_{\boldsymbol{\theta}, \mathbf{X}(\mathbf{e}_{\mathrm{vec}})} \right\rvert - \frac{M_N}{2}\log(2\pi) \\
    - \frac{1}{2} \left( \mathbf{p} - \mathbf{m}_{\boldsymbol{\theta}, \mathbf{X}(\mathbf{e}_{\mathrm{vec}})} \right)^\top \mathbf{C}_{\boldsymbol{\theta}, \mathbf{X}(\mathbf{e}_{\mathrm{vec}})}^{-1} \left( \mathbf{p} - \mathbf{m}_{\boldsymbol{\theta}, \mathbf{X}(\mathbf{e}_{\mathrm{vec}})}
    \right).
\end{multline}
To clarify the spatial information provided by the mean and covariance, define
\begin{equation}
    r_n \triangleq \left\lVert \mathbf{x}_n(\mathbf{e}_{\mathrm{vec}}) - \mathbf{x}_{\mathrm{Tx}} \right\rVert,
\end{equation}
and
\begin{equation}
    \rho_{mn} \triangleq \left\lVert \mathbf{x}_m(\mathbf{e}_{\mathrm{vec}}) - \mathbf{x}_n(\mathbf{e}_{\mathrm{vec}}) \right\rVert.
\end{equation}
Let $\delta_{a,b}$ denote the Kronecker delta.
For $r_n > 0$, the gradient of the $n$-th mean component with respect to the positioning error of sensor $i$ is
\begin{equation}
    \label{eq:mean_offset_gradient}
    \nabla_{\mathbf{e}^{(i)}} \left[ \mathbf{m}_{\boldsymbol{\theta}, \mathbf{X}(\mathbf{e}_{\mathrm{vec}})} \right]_n = \delta_{i, i_n} \frac{10\eta}{\ln 10} \frac{\mathbf{x}_n(\mathbf{e}_{\mathrm{vec}}) - \mathbf{x}_{\mathrm{Tx}}}{r_n(d_0 + r_n)}
\end{equation}
Thus, the mean provides spatial information along the radial direction from the transmitter to each corrected measurement location.
For $m \neq n$ with $\rho_{mn} > 0$, the corresponding gradient of the kernel entry is
\begin{multline}
    \label{eq:covariance_offset_gradient}
    \nabla_{\mathbf{e}^{(i)}} \left[ \mathbf{K}_{\boldsymbol{\theta},\mathbf{X}(\mathbf{e}_{\mathrm{vec}})} \right]_{m,n} \\
    = \frac{\ln 2}{d_{\mathrm{cor}}} \left[ \mathbf{K}_{\boldsymbol{\theta}, \mathbf{X}(\mathbf{e}_{\mathrm{vec}})} \right]_{m,n} \left( \delta_{i,i_m} - \delta_{i,i_n} \right) \frac{\mathbf{x}_m(\mathbf{e}_{\mathrm{vec}}) - \mathbf{x}_n(\mathbf{e}_{\mathrm{vec}})}{\rho_{mn}}
\end{multline}
Hence, the covariance provides information along pairwise directions between corrected measurement locations.
In particular, this gradient vanishes for pairs of measurements collected by the same sensor, because their common positioning offset does not change their pairwise distance.
Therefore, the covariance primarily constrains the relative offsets between different sensors, whereas the mean constrains the absolute placement of the corrected measurement locations through their distances from the known transmitter.
Consequently, when the measurement trajectories span only a limited range of directions relative to the transmitter, or when the directions of inter-sensor measurement pairs are insufficiently diverse, different positioning error vectors may produce similar mean vectors and covariance matrices. \par
To regularize these weakly identifiable directions, we adopt a MAP-based formulation for the positioning errors by incorporating the prior distribution introduced in Sec.\,\ref{sec:system_model}.
Because the positioning-error vectors are independent across sensors, their joint prior density is
\begin{equation}
    \label{eq:position_error_joint_prior}
    p_{\mathbf{e}}(\mathbf{e}_{\mathrm{vec}}) = \prod_{i=1}^{N} \phi \left( \mathbf{e}^{(i)}; \boldsymbol{\mu}_s, \boldsymbol{\Sigma}_s \right),
\end{equation}
where
\begin{multline}
    \label{eq:gaussian_pdf}
    \phi \left( \mathbf{x}; \boldsymbol{\mu}_s, \boldsymbol{\Sigma}_s \right) = \frac{1}{2\pi\sqrt{\left\lvert\boldsymbol{\Sigma}_s\right\rvert}} \\
    \times \exp \left[ - \frac{1}{2} \left( \mathbf{x} - \boldsymbol{\mu}_s \right)^\top \boldsymbol{\Sigma}_s^{-1} \left( \mathbf{x} - \boldsymbol{\mu}_s \right) \right].
\end{multline}
The prior also provides an absolute reference in the positioning-error space.
Specifically, its gradient with respect to the positioning error of sensor $i$ is
\begin{equation}
    \nabla_{\mathbf{e}^{(i)}} \log p_{\mathbf{e}}(\mathbf{e}_{\mathrm{vec}}) = -\boldsymbol{\Sigma}_s^{-1} \left( \mathbf{e}^{(i)} - \boldsymbol{\mu}_s \right).
\end{equation}
Thus, the prior anchors each sensor-specific positioning error around the prior mean $\boldsymbol{\mu}_s$ and regularizes directions that are weakly constrained by the RSS likelihood.
The resulting MAP-based objective is
\begin{align}
    \label{eq:MAP_compact}
    \mathcal{J} \left(\boldsymbol{\theta}, \mathbf{e}_{\mathrm{vec}} \right) &\triangleq \log \mathcal{L} \left( \mathbf{p} \mid \tilde{\mathbf{X}}, \boldsymbol{\theta}, \mathbf{e}_{\mathrm{vec}} \right) + \log p_{\mathbf{e}}(\mathbf{e}_{\mathrm{vec}}) \\
    \label{eq:MAP}
    &= \log \mathcal{L} \left( \mathbf{p} \mid \tilde{\mathbf{X}}, \boldsymbol{\theta}, \mathbf{e}_{\mathrm{vec}} \right) - \frac{N}{2} \log \left( 4\pi^2 \left\lvert \boldsymbol{\Sigma}_s \right\rvert \right) \notag\\
    &\quad - \frac{1}{2} \sum_{i=1}^{N} \left( \mathbf{e}^{(i)} - \boldsymbol{\mu}_s \right)^\top \boldsymbol{\Sigma}_s^{-1} \left( \mathbf{e}^{(i)} - \boldsymbol{\mu}_s \right).
\end{align}
The model parameters and positioning errors are jointly estimated by
solving
\begin{equation}
    \label{eq:joint_map_estimation}
    \left( \hat{\boldsymbol{\theta}}, \hat{\mathbf{e}}_{\mathrm{vec}} \right) = \underset{ \boldsymbol{\theta}\in\Theta,\;\mathbf{e}_{\mathrm{vec}}\in\mathcal{E}}{ \operatorname{arg\,max}} \; \mathcal{J} \left( \boldsymbol{\theta}, \mathbf{e}_{\mathrm{vec}} \right),
\end{equation}
where $\mathcal{E}$ denotes a finite box constraint set for the sensor-specific positioning errors.
Equation\,\eqref{eq:joint_map_estimation} corresponds to MAP estimation of the positioning errors jointly with maximum-likelihood estimation of $\boldsymbol{\theta}$ under the conditional model that treats the reported locations $\tilde{\mathbf{X}}$ as fixed.
Unlike the full generative posterior considered later in Sec.\,\ref{sec:performance_bounds}, this conditional formulation does not include the probability law of the true measurement trajectories.
Because the objective is generally nonconvex with respect to both $\boldsymbol{\theta}$ and $\mathbf{e}_{\mathrm{vec}}$, the numerical optimizer is not guaranteed to attain the global maximum.
In our implementation, we therefore obtain a numerical solution of Eq.\,\eqref{eq:joint_map_estimation} by minimizing the negative objective $-\mathcal{J}$ using the Adam optimizer\,\cite{kingmaAdamMethodStochastic2017},  with the constrained parameters reparameterized so that the optimization is performed in an unconstrained space.
The resulting predictor uses the estimated positioning errors and model parameters as plug-in estimates rather than marginalizing over their posterior uncertainty.

\subsection{Radio Map Construction}
Let $\hat{\mathbf{e}}^{(i)}$ denote the estimated positioning error of sensor $i$.
The corrected estimate of its $j$-th measurement location is
\begin{equation}
    \label{eq:corrected_position_estimate}
    \hat{\mathbf{x}}_j^{(i)} = \tilde{\mathbf{x}}_j^{(i)} - \hat{\mathbf{e}}^{(i)}.
\end{equation}
Using the global measurement index, let
\begin{align}
    \hat{\mathbf{e}}_n &\triangleq \hat{\mathbf{e}}^{(i_n)}, \\
    \hat{\mathbf{x}}_n &\triangleq \tilde{\mathbf{x}}_n-\hat{\mathbf{e}}_n
\end{align}
The corrected locations are collected into
\begin{equation}
    \label{eq:corrected_location_matrix_estimate}
    \hat{\mathbf{X}} \triangleq \left[ \hat{\mathbf{x}}_1, \cdots, \hat{\mathbf{x}}_{M_N} \right]^\top \in \mathbb{R}^{M_N \times 2}.
\end{equation}
Using the estimated parameters and corrected locations, the covariance matrix is
\begin{equation}
    \label{eq:position_error_aware_prediction_covariance}
    \mathbf{C}_{\hat{\boldsymbol{\theta}}, \hat{\mathbf{X}}} = \mathbf{K}_{\hat{\boldsymbol{\theta}}, \hat{\mathbf{X}}} + \hat{\sigma}_p^2 \mathbf{I}.
\end{equation}
The cross-covariance vector between $\mathbf{x}_\ast$ and the corrected measurement locations has entries
\begin{equation}
    \label{eq:position_error_aware_cross_covariance_vector}
    \left[ \mathbf{k}_{\hat{\boldsymbol{\theta}}, \hat{\mathbf{X}}}(\mathbf{x}_\ast) \right]_n = k_{\mathrm{exp},\hat{\boldsymbol{\theta}}} \left( \mathbf{x}_\ast, \hat{\mathbf{x}}_n \right), \quad n = 1, \cdots, M_N
\end{equation}
The RSS at $\mathbf{x}_\ast$ is estimated by the posterior mean
\begin{equation}
    \label{eq:position_error_aware_posterior_mean}
    \hat{p}(\mathbf{x}_\ast) = m_{\hat{\boldsymbol{\theta}}}(\mathbf{x}_\ast) + \mathbf{k}_{\hat{\boldsymbol{\theta}}, \hat{\mathbf{X}}}^\top(\mathbf{x}_\ast) \mathbf{C}_{\hat{\boldsymbol{\theta}}, \hat{\mathbf{X}}}^{-1} \left( \mathbf{p} - \mathbf{m}_{\hat{\boldsymbol{\theta}},\hat{\mathbf{X}}} \right).
\end{equation}
Evaluating Eq.\,\eqref{eq:position_error_aware_posterior_mean} over the query points yields the radio map.
The complete procedure for radio map construction is summarized in Alg.\,\ref{alg:position_error_aware}.
\begin{algorithm}[!t]
  \caption{Radio map construction and joint estimation of propagation
  parameters and position errors}
  \label{alg:position_error_aware}
  \begin{algorithmic}[1]
    \State \textbf{Input:}
    Dataset $\mathcal{D}$, transmitter location $\mathbf{x}_{\mathrm{Tx}}$, transmit power $P_{\mathrm{Tx}}$, error-prior mean $\boldsymbol{\mu}_s$, error-prior covariance $\boldsymbol{\Sigma}_s$, and query-point set $\mathcal{X}_{\ast}$
    \State \textbf{Output:}
    Radio map $\{\hat{p}(\mathbf{x}_{\ast})\}_{\mathbf{x}_{\ast}\in\mathcal{X}_{\ast}}$, estimated model parameters $\hat{\boldsymbol{\theta}}$, estimated positioning errors $\hat{\mathbf{e}}_{\mathrm{vec}}$, and corrected locations $\hat{\mathbf{X}}$
    \State Form $\tilde{\mathbf{X}}$ and $\mathbf{p}$ from $\mathcal{D}$
    \State Initialize $\boldsymbol{\theta}$ and $\mathbf{e}_{\mathrm{vec}}$
    \State Obtain $\hat{\boldsymbol{\theta}}$ and $\hat{\mathbf{e}}_{\mathrm{vec}}$ by numerically solving Eq.\,\eqref{eq:joint_map_estimation} using Adam to minimize $-\mathcal{J}(\boldsymbol{\theta},\mathbf{e}_{\mathrm{vec}})$
    \State Form the corrected-location matrix $\hat{\mathbf{X}}$ using Eq.\,\eqref{eq:corrected_location_matrix_estimate}
    \State Form $\mathbf{m}_{\hat{\boldsymbol{\theta}},\hat{\mathbf{X}}}$ using Eq.\,\eqref{eq:position_error_aware_mean_vector}
    \State Form $\mathbf{K}_{\hat{\boldsymbol{\theta}},\hat{\mathbf{X}}}$ using Eq.\,\eqref{eq:position_error_aware_kernel_matrix} and $\mathbf{C}_{\hat{\boldsymbol{\theta}},\hat{\mathbf{X}}}$ using Eq.\,\eqref{eq:position_error_aware_prediction_covariance}
    \ForAll{$\mathbf{x}_{\ast}\in\mathcal{X}_{\ast}$}
      \State Form $\mathbf{k}_{\hat{\boldsymbol{\theta}}, \hat{\mathbf{X}}}(\mathbf{x}_\ast)$ using Eq.\,\eqref{eq:position_error_aware_cross_covariance_vector}
      \State Compute $\hat{p}(\mathbf{x}_\ast)$ using Eq.\,\eqref{eq:position_error_aware_posterior_mean}
    \EndFor
    \State \Return $\{\hat{p}(\mathbf{x}_{\ast})\}_{\mathbf{x}_{\ast}\in\mathcal{X}_{\ast}}$, $\hat{\boldsymbol{\theta}}$, $\hat{\mathbf{e}}_{\mathrm{vec}}$, and $\hat{\mathbf{X}}$
  \end{algorithmic}
\end{algorithm}

\section{Performance Bound Analysis}
\label{sec:performance_bounds}
This section quantifies how closely a radio-map estimator approaches the minimum MSE achievable from the observed data.
We first define the conditional Bayes risk and decompose the estimator risk into irreducible and excess components.
Because the Bayes risk is generally intractable, we bound it using an oracle lower bound and a working-model upper bound, which in turn provide bounds on the excess risk.
The working-model analysis also clarifies the identifiability of trajectory-wise positioning offsets from trajectory information alone.

\subsection{Bayes Risk and Excess Risk}
Throughout this section, the propagation parameter vector $\boldsymbol{\theta}$ is not treated as a random variable, and the estimation performance is evaluated conditional on its true, fixed value.
Hereafter, $\mathbb{E}_{\boldsymbol{\theta}}[\cdot]$ denotes expectation under the generative model with $\boldsymbol{\theta}$ held fixed, where the expectation is taken over the true measurement trajectories, positioning errors, shadowing field, measurement noise, and, when applicable, the internal randomness of an estimator.
The evaluation-point set $\mathcal{X}_\ast = \{ \mathbf{x}_{\ast,1}, \cdots, \mathbf{x}_{\ast,M_\ast} \}$ is assumed to be deterministic, and the evaluation points themselves are free of positioning errors.
The true radio map values at the evaluation points are defined as
\begin{equation}
    \mathbf{f}_\ast \triangleq \left[ f(\mathbf{x}_{\ast,1}), \cdots, f(\mathbf{x}_{\ast,M_\ast}) \right]^\top \in \mathbb{R}^{M_\ast}.
\end{equation}
Let the matrix containing all true measurement locations be
\begin{equation}
    \mathbf{X} = \left[ \mathbf{x}_1, \cdots, \mathbf{x}_{M_N} \right]^\top \in \mathbb{R}^{M_N \times 2},
\end{equation}
where $\mathbf{X}$ is assumed to be a random variable following a known trajectory-generation law that is independent of the propagation environment.
Its joint probability density is denoted by $p_{\mathbf{X}}(\mathbf{X})$.
The joint prior density of the positioning-error vector $\mathbf{e}_{\mathrm{vec}}$ is given by
\begin{equation}
    p_{\mathbf{e}}(\mathbf{e}_{\mathrm{vec}}) \triangleq \prod_{i = 1}^N \phi\left( \mathbf{e}^{(i)}; \boldsymbol{\mu}_s, \boldsymbol{\Sigma}_s \right).
\end{equation}
For a given measurement-location matrix $\mathbf{X}$ and $\boldsymbol{\theta}$, let $p_{\mathrm{GP}}(\mathbf{p} \mid \mathbf{X}, \boldsymbol{\theta})$ denote the multivariate Gaussian probability density function of the RSS observation vector $\mathbf{p}$ induced by the GP observation model. \par
An arbitrary estimator constructed from the observed data $\mathcal{D}$ and an internal random variable $\mathbf{Z}_\delta$, which is independent of both $\mathcal{D}$ and the true generative process, is written as
\begin{equation}
    \label{eq:estimator_definition}
    \hat{\mathbf{f}}_\delta = \delta(\mathcal{D}, \mathbf{Z}_\delta).
\end{equation}
The risk of estimator $\delta$ is defined as the MSE per evaluation point:
\begin{equation}
    R_\delta(\boldsymbol{\theta}) \triangleq \frac{1}{M_\ast} \mathbb{E}_{\boldsymbol{\theta}}\left[ \left\lVert \mathbf{f}_\ast - \hat{\mathbf{f}}_\delta \right\rVert_2^2 \right].
\end{equation}
For fixed reported locations $\tilde{\mathbf{X}}$, the true measurement locations corresponding to a positioning-error vector $\mathbf{e}_{\mathrm{vec}}$ are determined by $\mathbf{X}(\mathbf{e}_{\mathrm{vec}})$ defined in Sec.\,\ref{sec:position_error_aware}.
Therefore, under the true generative model, the posterior density of the positioning errors is given by
\begin{equation}
    \pi_{\mathrm{true}}(\mathbf{e}_{\mathrm{vec}} \mid \mathcal{D}, \boldsymbol{\theta}) \propto p_{\mathrm{GP}}(\mathbf{p} \mid \mathbf{X}(\mathbf{e}_{\mathrm{vec}}), \boldsymbol{\theta})p_{\mathbf{e}}(\mathbf{e}_{\mathrm{vec}})p_{\mathbf{X}}(\mathbf{X}(\mathbf{e}_{\mathrm{vec}})).
\end{equation}
Thus, in addition to the likelihood of the RSS observations and the prior distribution of the positioning errors, the true posterior depends on how plausible the corrected measurement trajectory is under the trajectory-generation law.
For a fixed positioning-error vector $\mathbf{e}_{\mathrm{vec}}$, the GP posterior mean at the evaluation points is defined as
\begin{multline}
    \boldsymbol{\mu}_{\mathrm{GP}}(\mathbf{e}_{\mathrm{vec}}; \mathcal{D}, \boldsymbol{\theta}) = \\
    \mathbf{m}_{\boldsymbol{\theta}, \mathcal{X}_\ast} + \mathbf{K}_{\ast \mathbf{X}(\mathbf{e}_{\mathrm{vec}})} \mathbf{C}^{-1}_{\boldsymbol{\theta}, \mathbf{X}(\mathbf{e}_{\mathrm{vec}})} [\mathbf{p} - \mathbf{m}_{\boldsymbol{\theta}, \mathbf{X}(\mathbf{e}_{\mathrm{vec}})}].
\end{multline}
Under squared-error loss, the Bayes-optimal estimator conditional on the fixed value of $\boldsymbol{\theta}$ is
\begin{equation}
    \hat{\mathbf{f}}_{\mathrm{B}} \triangleq \mathbb{E}_{\boldsymbol{\theta}}[\mathbf{f}_\ast \mid \mathcal{D}],
\end{equation}
and, by the law of iterated expectations, it can be written as
\begin{equation}
    \hat{\mathbf{f}}_{\mathrm{B}} = \int \boldsymbol{\mu}_{\mathrm{GP}}(\mathbf{e}_{\mathrm{vec}};\mathcal{D}, \boldsymbol{\theta}) \pi_{\mathrm{true}}(\mathbf{e}_{\mathrm{vec}} \mid \mathcal{D}, \boldsymbol{\theta})d\mathbf{e}_{\mathrm{vec}}.
\end{equation}
The Bayes risk achieved by this estimator is defined as
\begin{align}
    L_{\mathrm{B}}(\boldsymbol{\theta}) &\triangleq \frac{1}{M_\ast} \mathbb{E}_{\boldsymbol{\theta}}\left[ \left\lVert \mathbf{f}_\ast - \hat{\mathbf{f}}_{\mathrm{B}} \right\rVert_2^2 \right] \\
    &= \frac{1}{M_\ast} \mathbb{E}_{\boldsymbol{\theta}}\left[ \operatorname{tr}\{ \operatorname{Cov}_{\boldsymbol{\theta}}(\mathbf{f}_\ast \mid \mathcal{D}) \} \right].
\end{align}
Thus, $L_{\mathrm{B}}$ is the minimum MSE achievable from the observed data under the assumed generative model.
Because $\boldsymbol{\theta}$ is fixed throughout the analysis, $L_{\mathrm{B}}(\boldsymbol{\theta})$ is a $\boldsymbol{\theta}$-conditional Bayes risk rather than a Bayes risk obtained by assigning a prior distribution to $\boldsymbol{\theta}$.
Since $\pi_{\mathrm{true}}$ contains the trajectory density $p_{\mathbf{X}}$ and the positioning errors affect both the GP mean and covariance nonlinearly, $\hat{\mathbf{f}}_{\mathrm{B}}$ and $L_{\mathrm{B}}$ are generally difficult to evaluate in closed form. \par
Because $\hat{\mathbf{f}}_{\mathrm{B}}$ is the conditional mean of $\mathbf{f}_\ast$ given $\mathcal{D}$, the orthogonality property of conditional expectation gives
\begin{equation}
    \mathbb{E}_{\boldsymbol{\theta}} \left[ (\mathbf{f}_\ast - \hat{\mathbf{f}}_{\mathrm{B}})^\top (\hat{\mathbf{f}}_{\mathrm{B}} - \hat{\mathbf{f}}_\delta) \right] = 0.
\end{equation}
Therefore, for any estimator $\delta$,
\begin{equation}
    \label{eq:risk_decomposition}
    R_\delta(\boldsymbol{\theta}) = L_{\mathrm{B}}(\boldsymbol{\theta}) + G_\delta(\boldsymbol{\theta}),
\end{equation}
where
\begin{equation}
    G_\delta(\boldsymbol{\theta}) \triangleq \frac{1}{M_\ast}\mathbb{E}_{\boldsymbol{\theta}}\left[ \left\lVert \hat{\mathbf{f}}_\delta - \hat{\mathbf{f}}_{\mathrm{B}} \right\rVert_2^2 \right] \geq 0
\end{equation}
is the excess risk of estimator $\delta$.
Thus, Eq.\,\eqref{eq:risk_decomposition} decomposes the total MSE into the Bayes risk $L_{\mathrm{B}}$, which is the irreducible component given the observed data, and the estimator-dependent excess risk $G_\delta$, which is the additional, in-principle avoidable component.
This excess risk includes additional losses relative to the conditional Bayes-optimal estimator, such as those arising from estimating the unknown propagation parameters from the observations and from model mismatch.

\subsection{Oracle Lower Bound}
Because $L_{\mathrm{B}}$ is generally intractable under the true generative model, we introduce an oracle estimator that can access to the true measurement locations $\mathbf{X}$ in addition to the observed data $\mathcal{D}$, and use its risk as a lower bound on $L_{\mathrm{B}}$.
The oracle estimator is defined as
\begin{equation}
    \hat{\mathbf{f}}_{\mathrm{or}} \triangleq \mathbb{E}_{\boldsymbol{\theta}} \left[ \mathbf{f}_\ast \mid \mathcal{D}, \mathbf{X} \right].
\end{equation}
Under the assumed independence relations, once $\mathbf{X}$ and the RSS observations $\mathbf{p}$ are given, additionally conditioning on the reported locations $\tilde{\mathbf{X}}$ does not change the conditional distribution of $\mathbf{f}_\ast$, and hence
\begin{equation}
    \hat{\mathbf{f}}_{\mathrm{or}} = \mathbb{E}_{\boldsymbol{\theta}}\left[ \mathbf{f}_\ast \mid \mathbf{X}, \mathbf{p} \right].
\end{equation}
For fixed $\boldsymbol{\theta}$ and $\mathbf{X}$, $\mathbf{f}_\ast$ and $\mathbf{p}$ are jointly Gaussian.
Define the kernel matrices among the evaluation points, between the evaluation and measurement points, and among the measurement points, respectively, as
\begin{align}
    \mathbf{K}_{\ast\ast} &\triangleq \left[ k_{\mathrm{exp},\boldsymbol{\theta}}(\mathbf{x}_{\ast,m}, \mathbf{x}_{\ast,n}) \right]_{m,n=1}^{M_\ast}, \\
    \mathbf{K}_{\ast \mathbf{X}} &\triangleq \left[ k_{\mathrm{exp},\boldsymbol{\theta}}(\mathbf{x}_{\ast,m}, \mathbf{x}_n) \right]_{\substack{m = 1, \cdots, M_\ast \\ n = 1, \cdots, M_N}}, \\
    \mathbf{K}_{\mathbf{XX}} &\triangleq \left[ k_{\mathrm{exp},\boldsymbol{\theta}}(\mathbf{x}_m, \mathbf{x}_n) \right]_{m,n=1}^{M_N}, \\
    \mathbf{K}_{\mathbf{X}\ast} &\triangleq \mathbf{K}_{\ast \mathbf{X}}^\top.
\end{align}
By Gaussian conditioning, the posterior covariance of the oracle estimator is given by
\begin{equation}
    \boldsymbol{\Sigma}_{\mathrm{or}}(\boldsymbol{\theta};\mathbf{X}) = \mathbf{K}_{\ast\ast} - \mathbf{K}_{\ast \mathbf{X}}(\mathbf{K}_{\mathbf{XX}} + \sigma_p^2 \mathbf{I})^{-1} \mathbf{K}_{\mathbf{X}\ast}.
\end{equation}
Therefore, the conditional MSE for a given true measurement-location matrix $\mathbf{X}$ is
\begin{equation}
    \ell_{\mathrm{or}}(\boldsymbol{\theta}; \mathbf{X}) \triangleq \frac{1}{M_\ast} \operatorname{tr}\left[ \boldsymbol{\Sigma}_{\mathrm{or}}(\boldsymbol{\theta};\mathbf{X}) \right].
\end{equation}
Averaging this conditional MSE under the generative model with $\boldsymbol{\theta}$ fixed at its true value gives the oracle risk as
\begin{equation}
    L_{\mathrm{or}}(\boldsymbol{\theta}) \triangleq \mathbb{E}_{\boldsymbol{\theta}}\left[ \ell_{\mathrm{or}}(\boldsymbol{\theta};\mathbf{X}) \right].
\end{equation}
Because $\hat{\mathbf{f}}_{\mathrm{or}}$ is a conditional-mean estimator with access to more information than $\mathcal{D}$ alone, the projection property of conditional expectation gives
\begin{equation}
    \label{eq:lower_bayes}
    L_{\mathrm{or}}(\boldsymbol{\theta}) \leq L_{\mathrm{B}}(\boldsymbol{\theta}).
\end{equation}
Thus, $L_{\mathrm{or}}$ provides a lower bound on the Bayes risk.

\subsection{Working-Model Upper Bound and Its Tightness}
\label{subsec:working_upper_bound}
The risk of any estimator constructed solely from the observable information $\mathcal{D}$ is no smaller than $L_{\mathrm{B}}$.
We therefore introduce a working-model estimator as a reference estimator for the analysis.
In the working-model, the reported locations $\tilde{\mathbf{X}}$ are treated as a fixed design, and the trajectory density $p_{\mathbf{X}}(\mathbf{X}(\mathbf{e}_{\mathrm{vec}}))$ appearing in the true posterior is not used.
Accordingly, the working-model posterior is defined as
\begin{equation}
    \label{eq:working_posterior}
    \pi_{\mathrm{work}}(\mathbf{e}_{\mathrm{vec}} \mid \mathcal{D}, \boldsymbol{\theta}) \propto p_{\mathrm{GP}}(\mathbf{p} \mid \mathbf{X}(\mathbf{e}_{\mathrm{vec}}), \boldsymbol{\theta})p_{\mathbf{e}}(\mathbf{e}_{\mathrm{vec}}).
\end{equation}
The predictor obtained by marginalizing the positioning errors under this posterior is defined as
\begin{equation}
    \hat{\mathbf{f}}_{\mathrm{work}} \triangleq \int \boldsymbol{\mu}_{\mathrm{GP}}(\mathbf{e}_{\mathrm{vec}};\mathcal{D}, \boldsymbol{\theta})\pi_{\mathrm{work}}(\mathbf{e}_{\mathrm{vec}} \mid \mathcal{D}, \boldsymbol{\theta})d\mathbf{e}_{\mathrm{vec}}.
\end{equation}
Because this integral generally cannot be evaluated in closed form, it is approximated numerically as described in Sec.\,\ref{sec:numerical_evaluation}.
For a fixed true value of $\boldsymbol{\theta}$, $\hat{\mathbf{f}}_{\mathrm{work}}$ is an estimator determined solely by $\mathcal{D}$.
Its risk evaluated under the true generative model is defined as
\begin{equation}
    U_{\mathrm{work}}(\boldsymbol{\theta}) \triangleq \frac{1}{M_\ast} \mathbb{E}_{\boldsymbol{\theta}}\left[ \left\lVert \mathbf{f}_\ast - \hat{\mathbf{f}}_{\mathrm{work}} \right\rVert_2^2 \right].
\end{equation}
Because $\hat{\mathbf{f}}_{\mathrm{B}}$ is the conditional-mean estimator based on $\mathcal{D}$,
\begin{equation}
    U_{\mathrm{work}}(\boldsymbol{\theta}) = L_{\mathrm{B}}(\boldsymbol{\theta}) + \frac{1}{M_\ast}\mathbb{E}_{\boldsymbol{\theta}}\left[\left\lVert \hat{\mathbf{f}}_{\mathrm{work}} - \hat{\mathbf{f}}_{\mathrm{B}} \right\rVert_2^2 \right].
\end{equation}
Therefore,
\begin{equation}
    \label{eq:upper_bayes}
    L_{\mathrm{B}}(\boldsymbol{\theta}) \leq U_{\mathrm{work}}(\boldsymbol{\theta}),
\end{equation}
and $U_{\mathrm{work}}$ provides an upper bound on the Bayes risk.
This upper-bound relation does not require the working model to coincide with the true generative model or $\hat{\mathbf{f}}_{\mathrm{work}}$ to be close to the true Bayes-optimal estimator. \par
Combining the true posterior defined above with the working-model posterior in Eq.\,\eqref{eq:working_posterior}, their density ratio can be written as
\begin{equation}
    \label{eq:true_work_posterior_ratio}
    \frac{\pi_{\mathrm{true}}(\mathbf{e}_{\mathrm{vec}} \mid \mathcal{D}, \boldsymbol{\theta})}{\pi_{\mathrm{work}}(\mathbf{e}_{\mathrm{vec}} \mid \mathcal{D}, \boldsymbol{\theta})} = c(\mathcal{D}, \boldsymbol{\theta}) p_{\mathbf{X}} \left( \mathbf{X}(\mathbf{e}_{\mathrm{vec}}) \right),
\end{equation}
where $c(\mathcal{D},\boldsymbol{\theta})$ is independent of $\mathbf{e}_{\mathrm{vec}}$.
Hence, the discrepancy between the Bayes-optimal and working-model predictors arises from the trajectory-density factor omitted in the working model.
In particular, if $p_{\mathbf{X}}(\mathbf{X}(\mathbf{e}_{\mathrm{vec}}))$ is constant for $\pi_{\mathrm{work}}$-almost every $\mathbf{e}_{\mathrm{vec}}$, then $\pi_{\mathrm{true}}=\pi_{\mathrm{work}}$, and consequently $\hat{\mathbf{f}}_{\mathrm{B}}=\hat{\mathbf{f}}_{\mathrm{work}}$ and $L_{\mathrm{B}}(\boldsymbol{\theta})=U_{\mathrm{work}}(\boldsymbol{\theta})$.
One sufficient condition is a sensor-wise translation-invariant trajectory law with no active boundary or other absolute-location constraints over the relevant support.
On a bounded domain, however, boundary effects can break this invariance; therefore, the equality is not assumed in the numerical evaluation. \par
The same translation invariance also clarifies the information available for estimating the sensor-specific constant offsets from the reported trajectories alone.
Without using the RSS observations, the posterior of the positioning-error vector conditioned on the reported locations satisfies
\begin{equation}
    \label{eq:trajectory_only_offset_posterior}
    p(\mathbf{e}_{\mathrm{vec}} \mid \tilde{\mathbf{X}}) \propto p_{\mathbf{e}}(\mathbf{e}_{\mathrm{vec}}) p_{\mathbf{X}} \left( \mathbf{X}(\mathbf{e}_{\mathrm{vec}}) \right).
\end{equation}
Thus, when the trajectory law is translation invariant over the relevant support, the reported trajectory does not update the prior distribution of the constant offset.
Absolute offset information must then be supplied by absolute-location constraints or by additional observations.
In the proposed method, the RSS observations provide such information through the distance-dependent mean and the inter-sensor spatial covariance structure described in Sec.\,\ref{sec:position_error_aware}.

\subsection{Bounds on the Bayes Risk and Excess Risk}
From Eq.\,\eqref{eq:lower_bayes} and Eq.\,\eqref{eq:upper_bayes},
\begin{equation}
    \label{eq:bayes_risk_bounds}
    L_{\mathrm{or}}(\boldsymbol{\theta}) \leq L_{\mathrm{B}}(\boldsymbol{\theta}) \leq U_{\mathrm{work}}(\boldsymbol{\theta})
\end{equation}
holds.
Therefore, since $G_\delta(\boldsymbol{\theta}) = R_\delta(\boldsymbol{\theta}) - L_{\mathrm{B}}(\boldsymbol{\theta})$ for any estimator $\delta$,
\begin{multline}
    \label{eq:excess_risk_bounds}
    \left(R_\delta(\boldsymbol{\theta}) - U_{\mathrm{work}}(\boldsymbol{\theta})\right)^+
    \leq G_\delta(\boldsymbol{\theta}) \leq R_\delta(\boldsymbol{\theta}) - L_{\mathrm{or}}(\boldsymbol{\theta})
\end{multline}
is obtained, where $(\cdot)^+$ denotes the positive part.
Thus, without directly evaluating the true Bayes-optimal estimator $\hat{\mathbf{f}}_{\mathrm{B}}$ or the Bayes risk $L_{\mathrm{B}}$, the deviation of an arbitrary estimator constructed from the observed data $\mathcal{D}$ from the Bayes-optimal performance can be bounded. \par
To specialize Eq.\,\eqref{eq:excess_risk_bounds} to the two estimators considered in this study, let $R_{\mathrm{prop}}(\boldsymbol{\theta})$ and $G_{\mathrm{prop}}(\boldsymbol{\theta})$ denote the risk and excess risk, respectively, of the proposed method in Sec.\,\ref{sec:position_error_aware}, and let $R_{\mathrm{base}}(\boldsymbol{\theta})$ and $G_{\mathrm{base}}(\boldsymbol{\theta})$ denote those of the position-error-agnostic baseline in Sec.\,\ref{sec:without_accounting}.
Applying Eq.\,\eqref{eq:excess_risk_bounds} to these two estimators yields their corresponding excess-risk intervals.
These intervals are evaluated in Sec.\,\ref{sec:numerical_evaluation}.

\section{Numerical Evaluation}
\label{sec:numerical_evaluation}
\begin{figure}[t!]
    \centering
    \includegraphics[width=\linewidth]{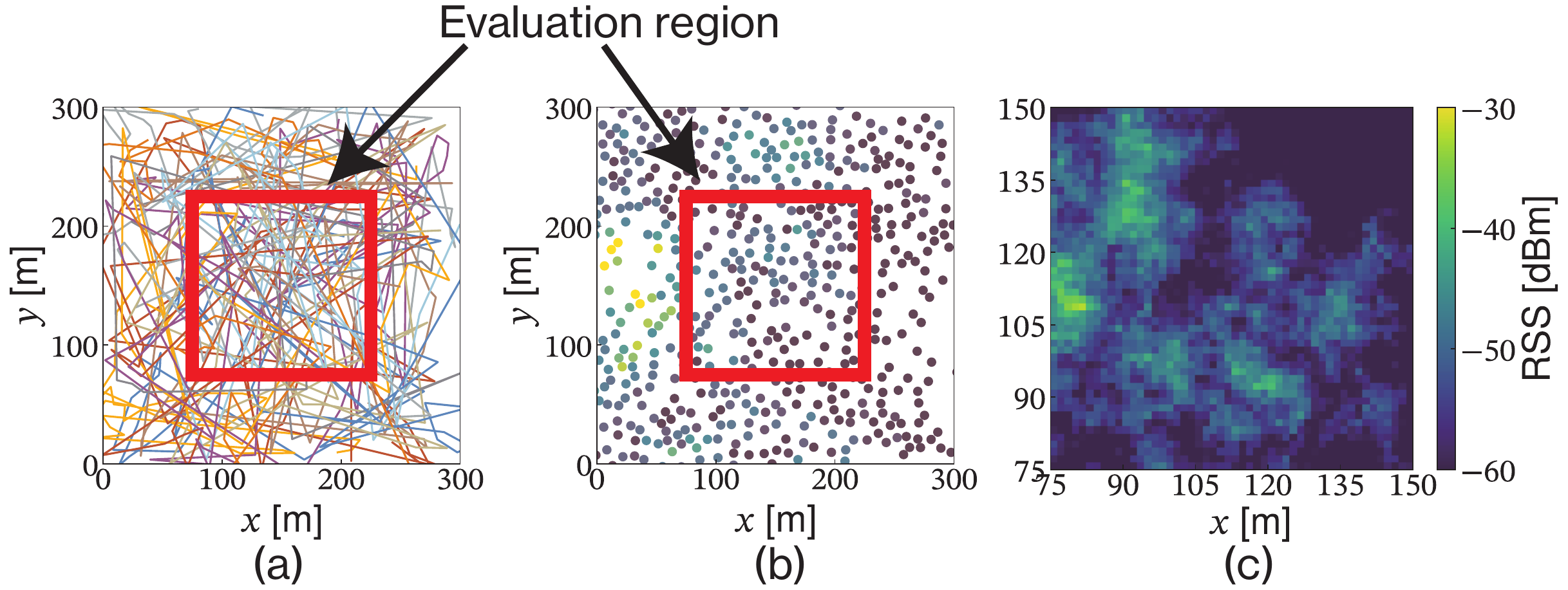}
    \caption{Example simulation: (a) sensor trajectories, (b) observed RSS, and (c) ground-truth radio map in the evaluated region.}
    \label{fig:example_of_simulation}
\end{figure}

This section evaluates the radio map estimation performance of the proposed method through 1000 independent numerical trials.
Using the Oracle lower bound and working-model upper bound derived in Sec.\,\ref{sec:performance_bounds}, we also evaluate the excess risks of the proposed method and the position-error-agnostic baseline.
The parameters listed in Table\,\ref{tab:reference_condition} are used as the reference condition\footnote{In practical scenarios, such as constructing public Wi-Fi radio maps, there are many cases where radio maps must be constructed for locations that are sufficiently far from the base station. Therefore, the base station was placed at $[0, 150]$, and the central region was designated as the evaluation region.}.
\begin{table}[t!]
    \centering
    \caption{Reference condition.}
    \begin{tabular}{c|c} \hline
        Parameter & Value \\ \hline
        Simulation area $\mathcal{A}$ & $300\times300\,[\mathrm{m}]$ \\
        Reference distance $d_0$ & $1.0\,[\mathrm{m}]$ \\
        BS location $\mathbf{x}_{\mathrm{Tx}}$ & $[0, 150]$ \\
        Transmit power $P_{\mathrm{Tx}}$ & $10\,[\mathrm{dBm}]$ \\
        Path-loss index $\eta$ & 3.0 \\
        Shadowing variance $\sigma_f^2$ & $64\,[\mathrm{dB}^2]$ \\
        Correlation distance $d_{\mathrm{cor}}$ & $20\,[\mathrm{m}]$ \\
        Measurement-noise variance $\sigma_p^2$ & $1.0\,[\mathrm{dB}^2]$ \\
        Number of sensors $N$ & 20 \\
        Measurement interval & $20\,[\mathrm{s}]$ \\
        Observation duration per sensor & $1800\,[\mathrm{s}]$ \\
        Learning rate of $\boldsymbol{\theta}$ & 0.24 \\
        Learning rate of $\mathbf{e}_{\mathrm{vec}}$ & 0.6 \\
        L\'evy walk parameter $\alpha$ & 0.5 \\
        L\'evy walk parameter $\beta$ & 1.0 \\ \hline
    \end{tabular}
    \label{tab:reference_condition}
\end{table}
The true measurement trajectories are generated using L\'evy walks\,\cite{rheeLevyWalkNatureHuman2011b}, and the sensor-specific quasi-static positioning errors are generated with $\boldsymbol{\mu}_s = 0$ and $\boldsymbol{\Sigma}_s = \operatorname{diag}[100, 100]$.
The MSE is evaluated on a $50\times50$ grid covering the central $150\,\mathrm{m}\times150\,\mathrm{m}$ region.
Before applying any of the estimation methods, the observations were sorted by sensor index and time.
To avoid numerical instability caused by closely spaced GP inputs, we apply global minimum-distance thinning with a threshold of 7.5 m.
Sensors with fewer than four retained measurements after thinning are excluded from radio-map construction\footnote{Since $7.5\,\mathrm{m}$ is sufficiently small relative to the $[300 \times 300]\,\mathrm{m}$ simulation domain, the effect of thinning is limited in practice.}.
Fig.\,\ref{fig:example_of_simulation} shows an example of the simulations. \par
To evaluate the dependence on propagation conditions, we conduct experiments under various settings in which the propagation parameters are varied from their reference values.
The proposed method, the position-error-agnostic baseline, NIGP, which treats input-location uncertainty as effective noise, and Ideal GPR, which uses the true measurement locations, are mainly considered in the evaluation. \par
The working-model estimator introduced in Sec.\,\ref{sec:performance_bounds} is numerically evaluated using 50 trials randomly selected from the 1000 trials for each propagation condition.
In the working model, the true propagation parameters $\boldsymbol{\theta}$ are assumed to be known, and the posterior $\pi_{\mathrm{work}}$ in Eq.\,\eqref{eq:working_posterior} is constructed by treating the reported locations as a fixed design.
Because $\pi_{\mathrm{work}}$ is a non-Gaussian distribution over $\mathbb{R}^{2N}$, the marginalization integral defining $\hat{\mathbf{f}}_{\mathrm{work}}$ cannot be evaluated in closed form.
We therefore introduce the tempering path from the positioning-error prior $p_{\mathbf{e}}$ to $\pi_{\mathrm{work}}$ as
\begin{equation}
    \label{eq:tempering_path}
    \pi_{\lambda}(\mathbf{e}_{\mathrm{vec}}) \propto
    p_{\mathrm{GP}}\left(\mathbf{p} \mid \mathbf{X}(\mathbf{e}_{\mathrm{vec}}),
    \boldsymbol{\theta}\right)^{\lambda}
    p_{\mathbf{e}}(\mathbf{e}_{\mathrm{vec}}),
    \quad \lambda \in [0, 1],
\end{equation}
and approximate $\hat{\mathbf{f}}_{\mathrm{work}}$ using an adaptive tempered sequential Monte Carlo (SMC) sampler\,\cite{nealAnnealedImportanceSampling2001,delmoralSequentialMonteCarlo2006}.
Since $\mathbf{e}_{\mathrm{vec}}$ is a fixed-dimensional latent variable rather than a time-series state, the particle population is updated along increasing values of $\lambda$ following the sequential-sampling framework for static models\,\cite{chopinSequentialParticleFilter2002}.
Each SMC replica uses $N_{\mathrm{p}} = 256$ particles.
The temperature sequence $\{\lambda_t\}$ is adaptively selected to target a conditional effective sample size of $0.9 N_{\mathrm{p}}$ and systematic resampling is performed whenever the ordinary effective sample size falls below $0.5 N_{\mathrm{p}}$
For each trial, eight independent replicas are executed, and their results are averaged to estimate $U_{\mathrm{work}}$.
The Oracle risk $L_{\mathrm{or}}$, the working-model risk $U_{\mathrm{work}}$, and the risks $R_{\mathrm{prop}}$ and $R_{\mathrm{base}}$ are all averaged over the same 50 trials, and each endpoint of the excess-risk intervals derived in Sec.\,\ref{sec:performance_bounds} is evaluated as a paired difference on these common trials.

\subsection{Radio Map Accuracy Performance}
\begin{figure}[t!]
    \centering
    \includegraphics[width=\linewidth]{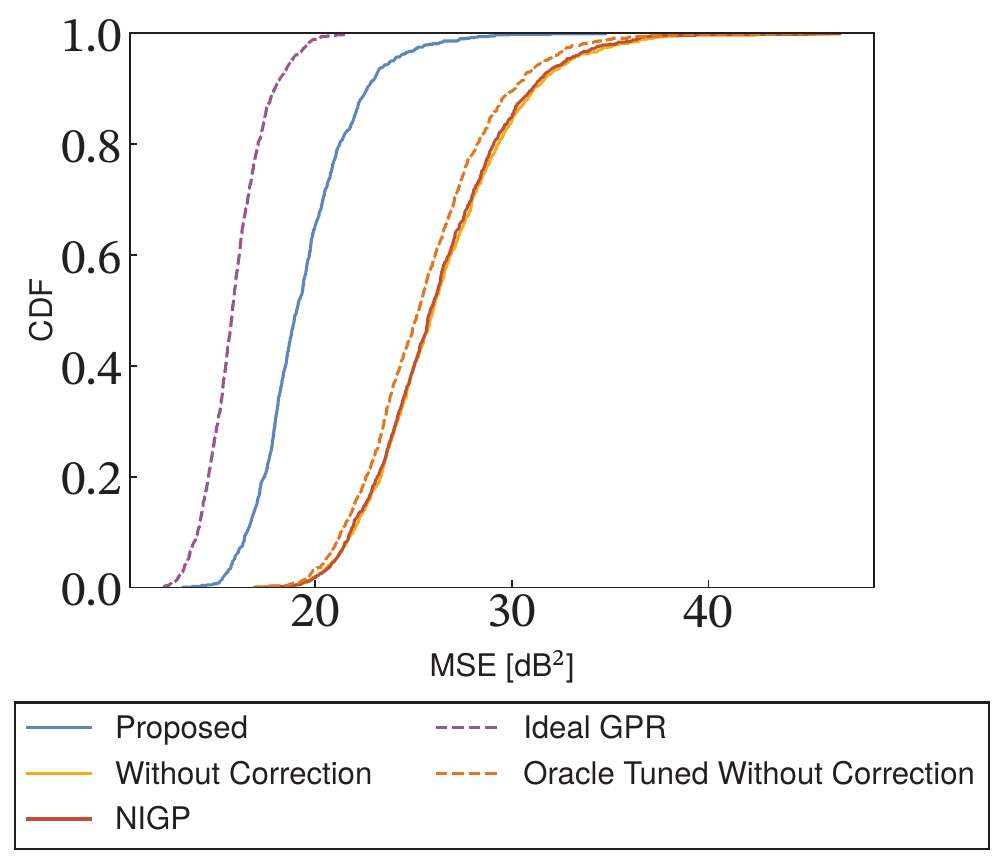}
    \caption{CDF of the radio-map estimation MSE under the reference condition.}
    \label{fig:baseline_cdf}
\end{figure}

The cumulative distribution function (CDF) of the radio map estimation MSE under the reference condition is shown in Fig.\,\ref{fig:baseline_cdf}.
\texttt{Ideal GPR} denotes radio map estimation using the true measurement locations, \texttt{Proposed} denotes the proposed method, \texttt{NIGP} denotes noisy-input GP, \texttt{Without Correction} denotes the position-error-agnostic baseline, and \texttt{Oracle Tuned Without Correction} denotes the same position-error-agnostic GP model whose objective is directly chosen to minimize the MSE against the true radio map\footnote{This method has access to the true radio map and therefore involves data leakage. Although it cannot be implemented in practice, it is included as a reference for assessing the representational limitation of the position-error-agnostic model.}.
Compared with \texttt{Ideal GPR}, the proposed method exhibits an increase in MSE of approximately $3.26\,\mathrm{dB}^2$, whereas \texttt{Without Correction} and NIGP exhibit increases of approximately $10.0\,\mathrm{dB}^2$.
Even \texttt{Oracle Tuned Without Correction}, which is allowed to exploit data leakage, exhibits an increase of approximately $9.47\,\mathrm{dB}^2$, leaving a clear gap from the proposed method.
These results demonstrate the effectiveness of the proposed method under the reference condition. \par
\begin{figure*}[!t]
    \centering
    \includegraphics[width=\linewidth]{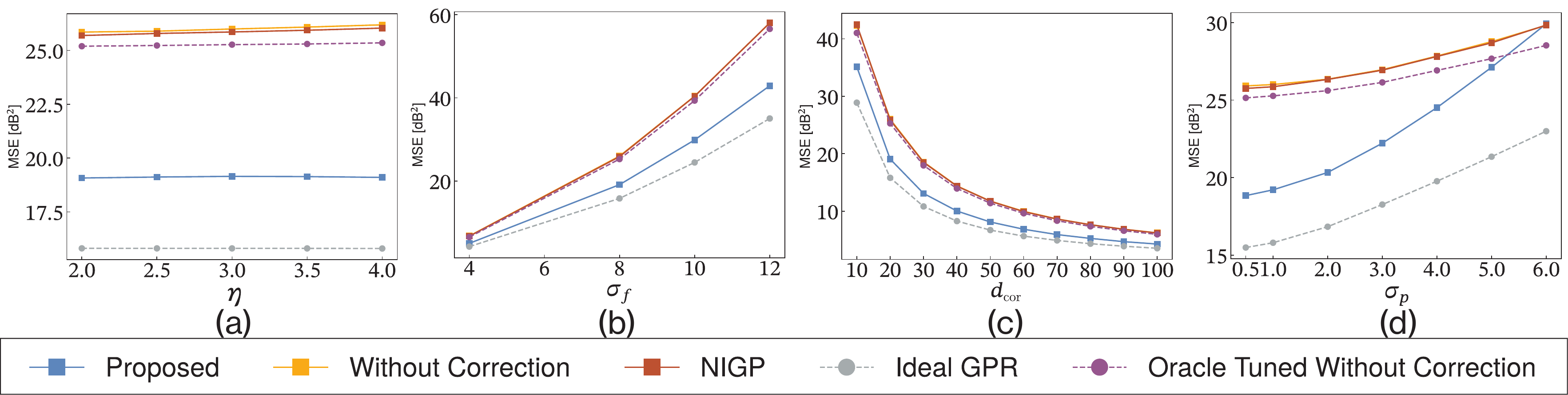}
    \caption{Effects of radio-propagation parameters on the median MSE: (a) $\eta$, (b) $\sigma_f$, (c) $d_{\mathrm{cor}}$, and (d) $\sigma_p$.}
    \label{fig:sweep_result}
\end{figure*}

Next, Fig.\,\ref{fig:sweep_result} shows the median MSE when $\eta$, $\sigma_f$, $d_{\mathrm{cor}}$, and $\sigma_p$ are varied from the reference condition.
The proposed method consistently outperformed \texttt{Without Correction} over almost all evaluated conditions.
In Fig.\,\ref{fig:sweep_result}(a), the performance of the proposed method is relatively insensitive to $\eta$.
The sensitivity of the distance-dependent mean to a positioning offset is proportional to $\eta$; hence, for a small $\eta$, the mean provides weaker absolute-location information and its variation is less distinguishable from stochastic shadowing.
The relatively stable performance of the proposed method therefore indicates that the pairwise spatial covariance provides complementary information for location calibration when the mean structure is less informative. \par
In Fig.\,\ref{fig:sweep_result}(b), the MSE increases with $\sigma_f$ even for \texttt{Ideal GPR}, reflecting the increased intrinsic variation of the shadowing field.
A larger shadowing amplitude also increases the effect of spatial misregistration, which explains the substantial degradation of the methods that do not explicitly correct the measurement locations. \par
As shown in Fig.\,\ref{fig:sweep_result}(c), increasing $d_{\mathrm{cor}}$ improves the performance of all methods because shadowing observations remain informative over a wider spatial range.
Moreover, a larger $d_{\mathrm{cor}}$ increases the number of inter-sensor measurement pairs with non-negligible covariance, providing additional pairwise constraints on the relative sensor offsets exploited by the proposed method. \par
Finally, in Fig.\,\ref{fig:sweep_result}(d), increasing $\sigma_p$ increases the diagonal measurement-noise term $\sigma_p^2 \mathbf{I}$ in the observation covariance matrix.
When this term dominates the spatial covariance, the RSS observations become less informative about both the distance-dependent mean and the pairwise covariance structure used for position calibration.
Consequently, the advantage of the proposed method gradually decreases with measurement noise and almost vanishes at $\sigma_p = 6\, \mathrm{dB}$ (i.e., $\sigma_p^2 = 36\,\mathrm{dB}^2$).
Overall, these results indicate that jointly exploiting the mean and spatial covariance structures enables effective position calibration over a broad range of propagation conditions, provided that the RSS measurements retain sufficient spatial information.

\subsection{Deviation from the Bayes-Optimal Performance}
\begin{figure*}
    \centering
    \includegraphics[width=\linewidth]{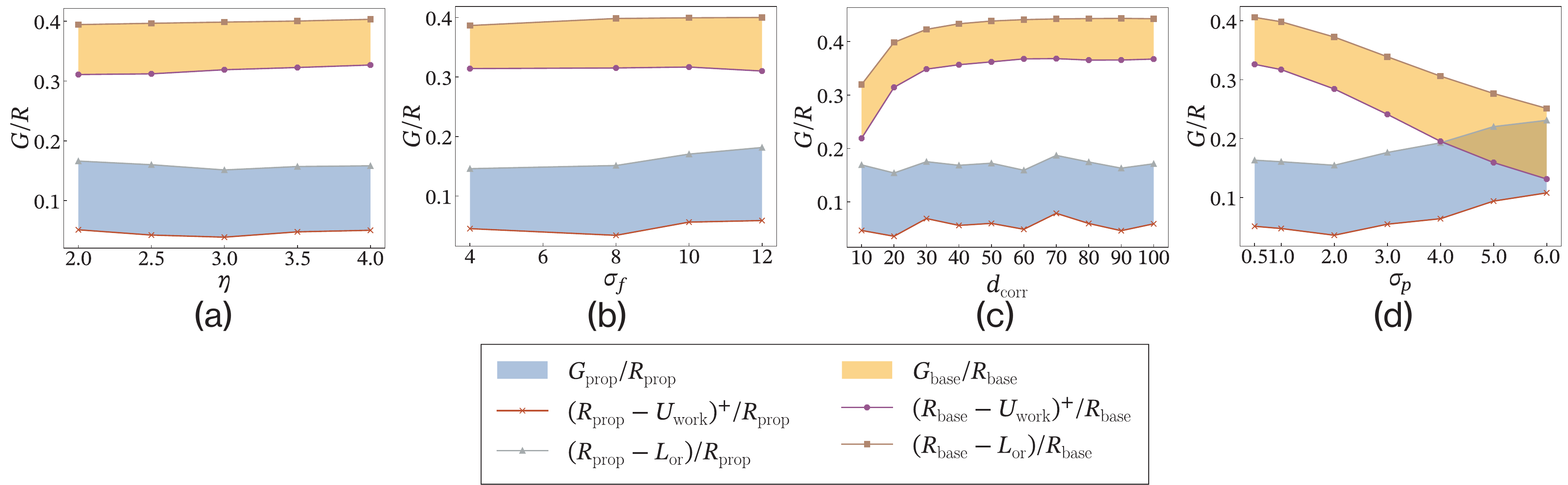}
    \caption{Effects of radio propagation parameters on the excess risk ratio: (a) $\eta$, (b) $\sigma_f$, (c) $d_{\mathrm{cor}}$, and (d) $\sigma_p$.}
    \label{fig:excess_risk}
\end{figure*}

Fig.\,\ref{fig:excess_risk} complements the absolute-MSE results in Fig.\,\ref{fig:sweep_result} by quantifying what fraction of each method's MSE is attributable to excess risk, i.e., error that could in principle be reduced by a Bayes-optimal use of the same observations.
Under the reference condition (Fig.\,\ref{fig:excess_risk}(a), $\eta=3$), the bound interval for $G_{\mathrm{prop}} / R_{\mathrm{prop}}$ is approximately 5--16\%, whereas that of $G_{\mathrm{base}} / R_{\mathrm{base}}$ is approximately 32--40\%.
Thus, only a relatively small fraction of the proposed method's MSE is attributable to estimator suboptimality, whereas roughly one third or more of the baseline MSE remains excess error.
This result indicates that the proposed method not only achieves a lower absolute MSE but also exploits the information contained in the observations substantially more effectively. \par
To interpret the ordinate, recall that the total risk of an estimator $\delta$ is decomposed as $R_{\delta} = L_{\mathrm{B}} + G_{\delta}$.
Here, $L_{\mathrm{B}}$ is the irreducible error remaining even when the observed data  are used optimally under the assumed model, whereas $G_{\delta}$ is the additional, in-principle avoidable error caused by deviation from the Bayes-optimal estimator.
Hence, $G_{\delta} / R_{\delta}$ represents the fraction of the total MSE attributable to this avoidable component, and a smaller value indicates performance closer to the Bayes-optimal benchmark.
Because $L_{\mathrm{B}}$ cannot be evaluated directly, we use Eq.\,\eqref{eq:excess_risk_bounds}: the lower and upper bounds of $G_{\delta} / R_{\delta}$ are $(R_\delta-U_{\mathrm{work}})^+/R_\delta$ and $(R_\delta-L_{\mathrm{or}})/R_\delta$, respectively.
Accordingly, the shaded regions in Fig.\,\ref{fig:excess_risk} represent bound intervals rather than statistical confidence intervals. \par
The normalized excess risk of the proposed method remains relatively stable with $\eta$ in Fig.\,\ref{fig:excess_risk}(a), indicating that the spatial covariance provides complementary information when the distance-dependent mean is less informative at small $\eta$.
Similarly, in Fig.\,\ref{fig:excess_risk}(b), the normalized excess-risk intervals change little with $\sigma_f$, even though the absolute MSE increases substantially, showing that larger shadowing variance primarily changes the difficulty of the estimation problem rather than the relative efficiency of the estimators.
In Fig.\,\ref{fig:excess_risk}(c), the difference between the two methods becomes larger with $d_{\mathrm{cor}}$ because the wider correlation range provides more inter-sensor measurement pairs that carry useful information about relative positioning offsets, which is exploited by the proposed method.
Finally, in Fig.\,\ref{fig:excess_risk}(d), the two intervals converge as $\sigma_p$ increases.
This convergence does not indicate improved absolute accuracy of the baseline; rather, strong measurement noise both suppresses the spatial information available for position calibration and increases the irreducible component of the estimation error.
Consequently, the benefit of position correction becomes negligible in the noise-dominated regime. \par
Overall, Figs.\,\ref{fig:sweep_result} and \ref{fig:excess_risk} together show that the proposed method is particularly effective when the RSS observations retain informative spatial structure: it achieves both a lower total MSE and a substantially smaller fraction of avoidable error than the position-error-agnostic baseline.

\section{Evaluation Based on Measured GNSS Error Models}
\label{sec:real_gnss_evaluation}
This section evaluates the robustness of the proposed method to temporally correlated positioning-error components that are not explicitly represented by the sensor-specific constant-offset model.
To construct measurement-informed positioning-error processes, we identify AR(2) models from GNSS positioning errors measured with two smartphones and use the resulting models to generate GNSS-based observation trajectories.
Because the numerical experiments in Sec.\,\ref{sec:numerical_evaluation} use a 20\,s measurement interval, the AR(2) models identified from the 1\,Hz measurements are converted to approximate 20\,s models before simulation.
We also compare the proposed method with a KF--RTS trajectory-smoothing baseline that explicitly exploits the temporal correlation of the positioning errors.
Details of the GNSS measurements and model validation, the conversion of the 1\,Hz models to model AR(2) models sampled at 20\,s intervals, and the complete state-space formulation of the KF--RTS baseline are given in Appendices\,\ref{app:gnss_model_details} and\,\ref{app:kf_rts_details}.
Note that the RSS observations and true trajectories remain simulated; the measured GNSS data are used to construct the positioning-error models.

\subsection{GNSS-Derived Error Model and Evaluation Setup}
\label{subsec:gnss_evaluation_setup}
Let $\mathbf{x}_{\mathrm{ref},k}\in\mathbb{R}^2$ and $\mathbf{x}_{\mathrm{obs},k}\in\mathbb{R}^2$ denote the reference and smartphone-reported GNSS positions, respectively, at the 1\,Hz time index $k$.
The measured positioning error is decomposed into a quasi-static component $\mathbf{a}$ and a temporally varying residual $\mathbf{u}_k$ as
\begin{equation}
    \label{eq:gnss_error_decomposition}
    \mathbf{e}_{\mathrm{obs},k} \triangleq \mathbf{x}_{\mathrm{obs},k}-\mathbf{x}_{\mathrm{ref},k} = \mathbf{a}+\mathbf{u}_k.
\end{equation}
For each axis $q\in\{x,y\}$, the residual is modeled by an AR(2) process,
\begin{equation}
    \label{eq:gnss_ar2_model}
    u_{q,k} = \omega_{q,1}u_{q,k-1} + \omega_{q,2}u_{q,k-2} + \epsilon_{q,k}, \qquad \epsilon_{q,k}\sim\mathcal{N}(0,\sigma_q^2).
\end{equation}
The quasi-static component $\mathbf{a}$ corresponds to the sensor-specific positioning offset assumed by the proposed method, whereas $\mathbf{u}_k$ represents the time-varying component that is not explicitly modeled by the proposed method. \par
The error models were identified from GNSS measurements collected using an RTK reference receiver and two smartphones, a motorola edge 40 neo and a moto g24.
The detailed measurement configuration, observed trajectories, identified AR(2) parameters, and validation results are reported in Appendix\,\ref{app:gnss_model_details}.
Although the fitted AR(2) models do not reproduce the complete marginal distributions of the measured errors, they approximately reproduce the strength and decay characteristics of the measured autocorrelation, which are the temporal characteristics of interest in this evaluation. \par
To make the identified error models consistent with the 20\,s measurement interval used in Sec.\,\ref{sec:numerical_evaluation}, each 1\,Hz AR(2) model is approximated by a 20\,s AR(2) model that matches the stationary autocovariances at lags 0, 20, and 40\,s.
The corresponding Yule--Walker construction is given in Appendix\,\ref{app:gnss_model_details}. \par
As in Sec.\,\ref{sec:numerical_evaluation}, the true trajectories of $N=20$ sensors are generated independently in each trial.
Because the measured values of $\mathbf{a}$ are particular realizations associated with the individual devices, they are not assigned identically to all simulated sensors.
Instead, the quasi-static positioning bias of sensor $i$ is generated as
\begin{equation}
    \label{eq:ar2_simulation_bias}
    \mathbf{b}^{(i)} \sim \mathcal{N}\left(\mathbf{0},100\mathbf{I}\right).
\end{equation}
Let $\bar{\mathbf{u}}_n^{(i)}$ denote an independent residual sequence generated from the corresponding 20\,s AR(2) model.
The observed position is then generated as
\begin{equation}
    \label{eq:ar2_simulated_observed_position}
    \tilde{\mathbf{x}}_n^{(i)} = \mathbf{x}_n^{(i)} + \mathbf{b}^{(i)} + \overline{\mathbf{u}}_n^{(i)}.
\end{equation}
The AR(2) states are initialized from their stationary distributions.
The prior distribution of the constant positioning bias used by the proposed method is also set to $\mathcal{N}(\mathbf{0},100\mathbf{I})$.
Thus, the quasi-static component is statistically matched between the data-generation model and the proposed method, whereas the additional time-varying AR(2) residual constitutes deliberate positioning-error model mismatch. \par
For comparison, we use a KF--RTS baseline whose linear Gaussian state-space model includes the sensor position and velocity, a sensor-specific constant bias, and the AR(2) residual.
The baseline is provided with the same 20\,s AR(2) coefficients and innovation variances as those used to generate the positioning errors, but the true positions, realized biases, and realized residuals are not provided.
Only the GNSS observation trajectories are used in the KF--RTS smoothing stage; the RSS observations are introduced only in the subsequent GPR stage. The complete state-space model and initialization are given in Appendix\,\ref{app:kf_rts_details}.

\subsection{Radio Map Estimation Performance under Temporally Correlated Positioning Errors}
\label{subsec:ar2_radio_map_evaluation}
\begin{figure}[t!]
    \centering
    \includegraphics[width=\linewidth]{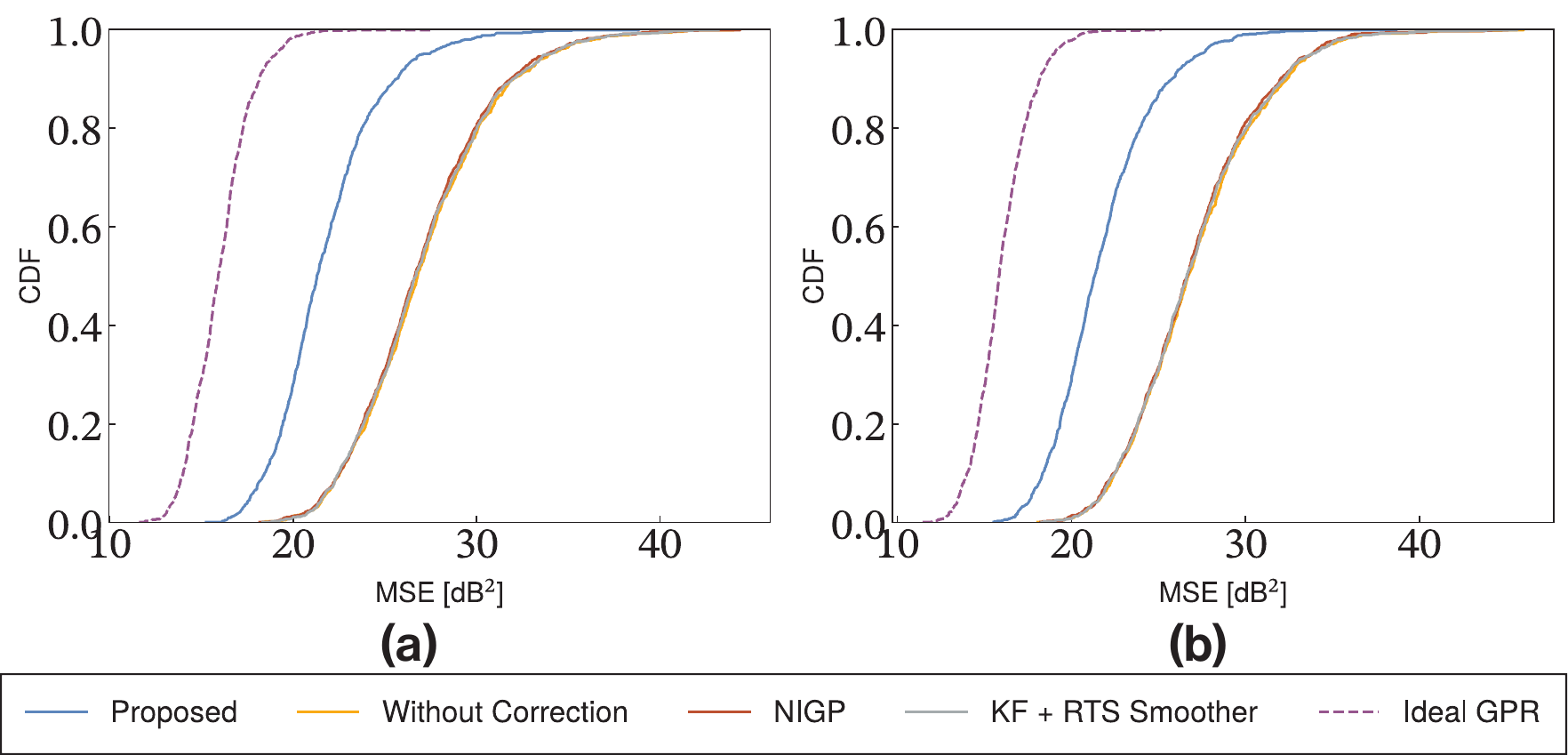}
    \caption{CDF of radio-map MSE with time-correlated positioning errors:
    (a) moto g24 and (b) motorola edge 40 neo.}
    \label{fig:AR2_CDF}
\end{figure}

Using the setup in Sec.\,\ref{subsec:gnss_evaluation_setup}, we conduct 1000 independent trials for each device under the same propagation parameters and true trajectory-generation conditions as the reference condition in Sec.\,\ref{sec:numerical_evaluation}.
Because the proposed method estimates only a sensor-specific constant positioning bias, the time-varying AR(2) residual in Eq.\,\eqref{eq:ar2_simulated_observed_position} constitutes deliberate model mismatch.
The CDFs of the radio map MSE obtained by the different methods are shown in Fig.\,\ref{fig:AR2_CDF}.
As shown in Fig.\,\ref{fig:AR2_CDF}, although \texttt{KF + RTS Smoother} is provided with the AR(2) dynamics used to generate the positioning errors, its radio map estimation performance under the present evaluation conditions is close to that of \texttt{Without Correction}.
This behavior is consistent with the translation ambiguity described in Sec.\,\ref{subsec:working_upper_bound} and formalized by the state-space model in Appendix\,\ref{app:kf_rts_details}.
The KF + RTS smoother can exploit the temporal correlation to estimate the time-varying AR(2) residual, but the GNSS trajectory likelihood cannot separately identify a sensor-specific quasi-static bias from a constant translation of the true trajectory.
Consequently, the separation of these two components is determined only by the absolute-position and bias priors, and a residual sensor-wise translation can remain in the smoothed trajectory.
Since the resulting positions are subsequently used as GPR inputs, this residual spatial misregistration limits the improvement over \texttt{Without Correction}. \par
In contrast, although the proposed method does not explicitly model the time-varying residual, it achieves lower MSE than both \texttt{Without Correction} and \texttt{KF + RTS Smoother} for the error models of both devices.
The proposed method uses spatial information from the RSS observations that is unavailable to the trajectory-only smoother: the distance-dependent mean provides information about absolute placement relative to the known BS, while the inter-sensor covariance provides information about relative sensor offsets.
These RSS-derived constraints provide additional information for estimating the quasi-static positioning offsets even in the presence of the unmodeled time-varying residual.
At the median of the CDF, the MSE of the proposed method increases by only approximately $5\,\mathrm{dB}^2$ relative to \texttt{Ideal GPR}, whereas the other non-ideal methods exhibit increases of approximately $10\,\mathrm{dB}^2$ or more.
These results indicate that, under the trajectory-generation conditions and temporal-correlation models considered here, modeling the temporal structure of the positioning errors alone is insufficient to reliably remove the sensor-specific quasi-static displacement, whereas incorporating the spatial structure of the RSS observations provides effective additional information for post-hoc location calibration. \par
Importantly, the GNSS trajectory likelihood of this baseline is invariant to a sensor-wise transformation $\mathbf{p}_n^{(i)}\mapsto\mathbf{p}_n^{(i)}+\mathbf{c}^{(i)}$ and $\mathbf{b}^{(i)}\mapsto\mathbf{b}^{(i)}-\mathbf{c}^{(i)}$ for any constant $\mathbf{c}^{(i)}\in\mathbb{R}^2$.
Hence, except for the absolute-location information provided by the initial-position and bias priors, the GNSS trajectory alone cannot distinguish a constant translation of the trajectory from the corresponding constant positioning bias.
This structural ambiguity is consistent with the trajectory-only identifiability discussion in Sec.\,\ref{subsec:working_upper_bound}.

\section{Conclusion}
\label{sec:conclusion}
This paper addressed radio map construction in the presence of temporally persistent, sensor-specific positioning errors.
We developed a GPR-based radio map construction method with post-hoc location calibration, which jointly estimates sensor-specific position offsets and radio propagation parameters from the  RSS observations and constructs the radio map using the calibrated measurement locations.
By exploiting both the distance-dependent path-loss structure and the spatial correlation of shadowing, the proposed method uses the radio measurements themselves as additional information for correcting the locations at which they were collected. \par
Numerical evaluations demonstrated the effectiveness of the proposed method over a wide range of propagation conditions.
Under the reference condition, the proposed method exhibited an MSE gap of approximately $3.26\,\mathrm{dB}^2$ relative to Ideal GPR, whereas the position-error-agnostic and NIGP methods exhibited gaps of approximately $10\,\mathrm{dB}^2$.
The benefit of location calibration was particularly pronounced in low-to-moderate measurement-noise regimes, where the RSS observations retain sufficient spatial information for estimating the positioning offsets.
Further, results of the theoretical analysis based on a conditional MMSE benchmark showed that the fraction of MSE attributable to excess risk was approximately 24--27 percentage points lower for the proposed method than for the position-error-agnostic baseline. \par
Finally, we evaluated robustness to positioning-error model mismatch using GNSS-based error sequences derived from measured GNSS data.
Although the proposed method assumes a constant positioning offset for each sensor and does not explicitly model the remaining time-varying error component, it consistently outperformed both the position-error-agnostic baseline and the KF + RTS smoother baseline for the GNSS error models considered.
These results demonstrate that exploiting the spatial structure of RSS for both radio-map construction and post-hoc location calibration is effective under temporally correlated positioning errors.
\par
An interesting direction for future work is to incorporate absolute-location constraints, such as road networks, walkable regions, or known anchors, through $p_{\mathbf{X}}$ and the corresponding trajectory-density term in the calibration objective, thereby breaking the translation symmetry and providing additional information about the sensor-specific offsets.

\appendices

\section{Details of the GNSS Error Model and Its 20\,s Approximation}
\label{app:gnss_model_details}
This appendix details the GNSS measurements and the construction of the 20-s AR(2) error models used in Sec.\,\ref{sec:real_gnss_evaluation}.
We first describe the GNSS measurement procedure and validate the AR(2) models identified from the measured positioning errors.
Then, the construction of the approximate 20\,s AR(2) models, used in the numerical evaluation, is described.

\subsection{GNSS Measurements and AR(2) Model Validation}
\label{app:gnss_measurement_validation}
\begin{figure}[t!]
    \centering
    \includegraphics[width=0.9\linewidth]{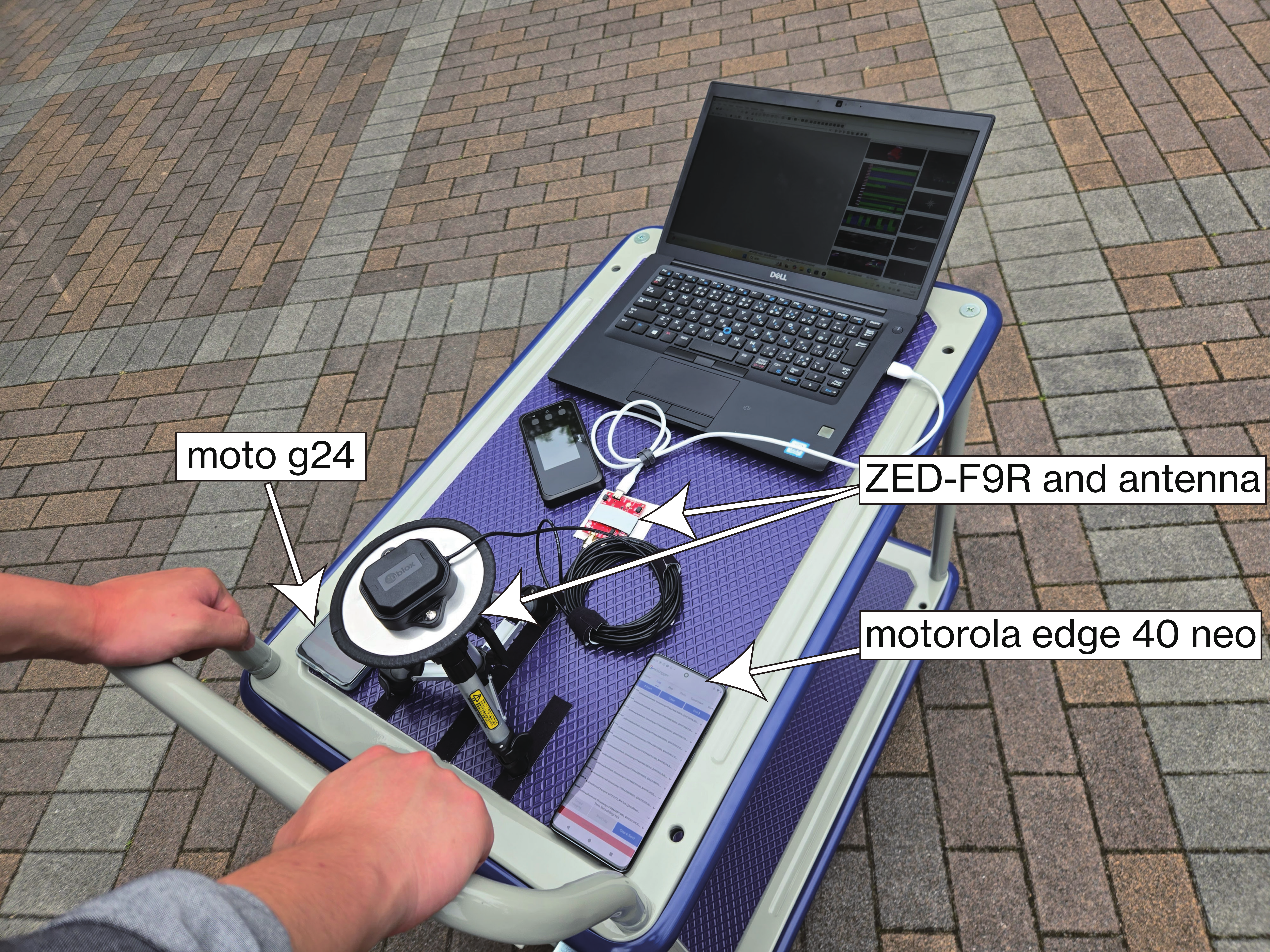}
    \caption{Measurement setup.}
    \label{fig:observations}
\end{figure}

\begin{figure}[t!]
    \centering
    \includegraphics[width=0.9\linewidth]{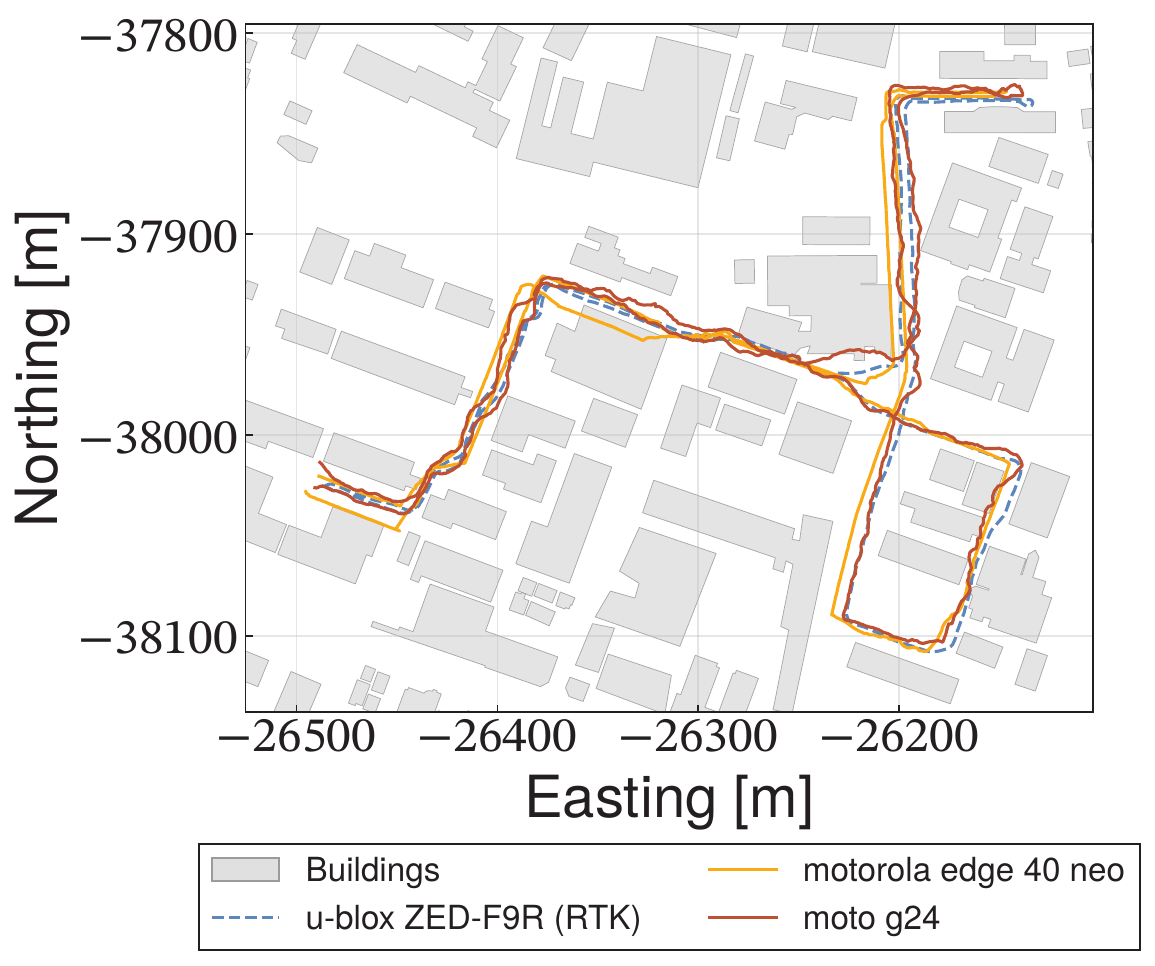}
    \caption{Observed trajectory
    (the building data is licensed under \copyright OpenStreetMap contributors).}
    \label{fig:observation_trajectory}
\end{figure}

\begin{figure}[t!]
    \centering
    \subfloat[]{\includegraphics[width=0.45\linewidth]{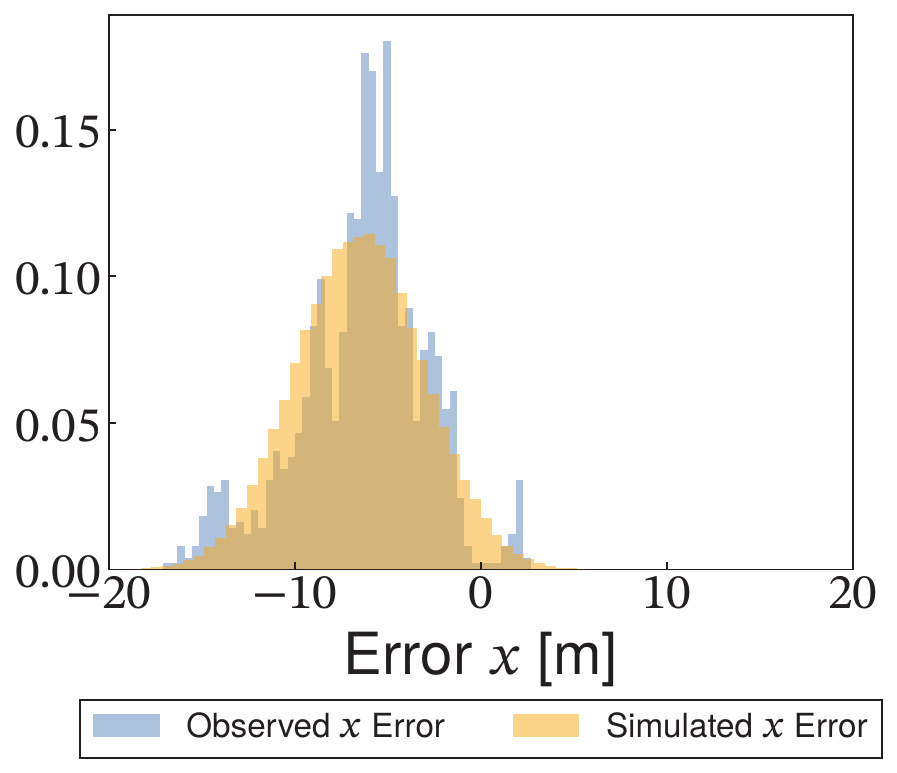}}
    \subfloat[]{\includegraphics[width=0.45\linewidth]{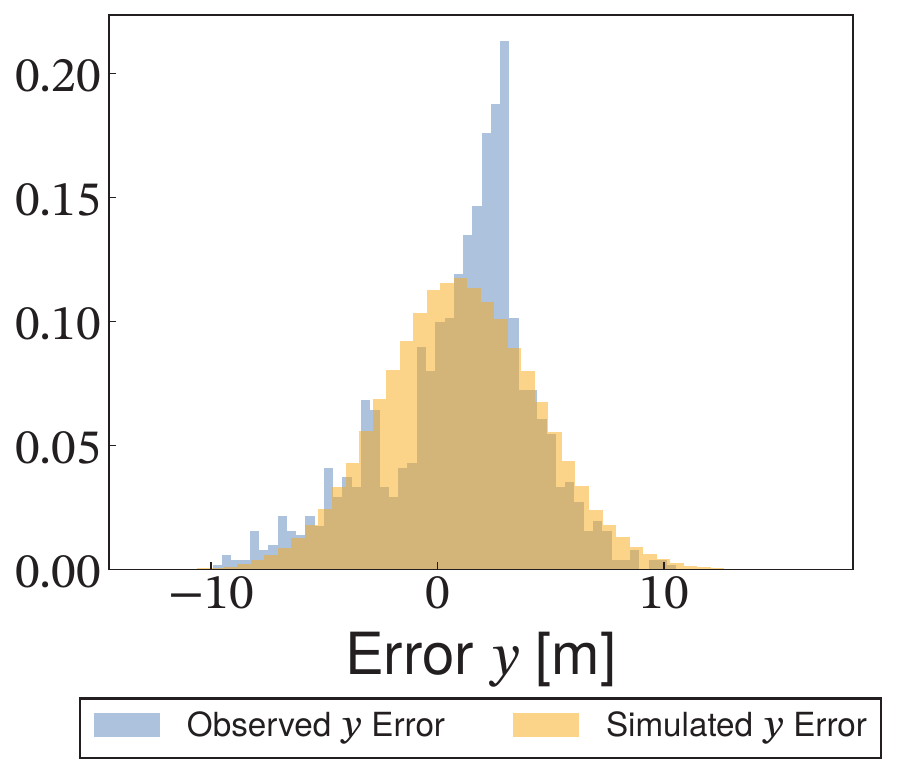}} \\
    \subfloat[]{\includegraphics[width=0.45\linewidth]{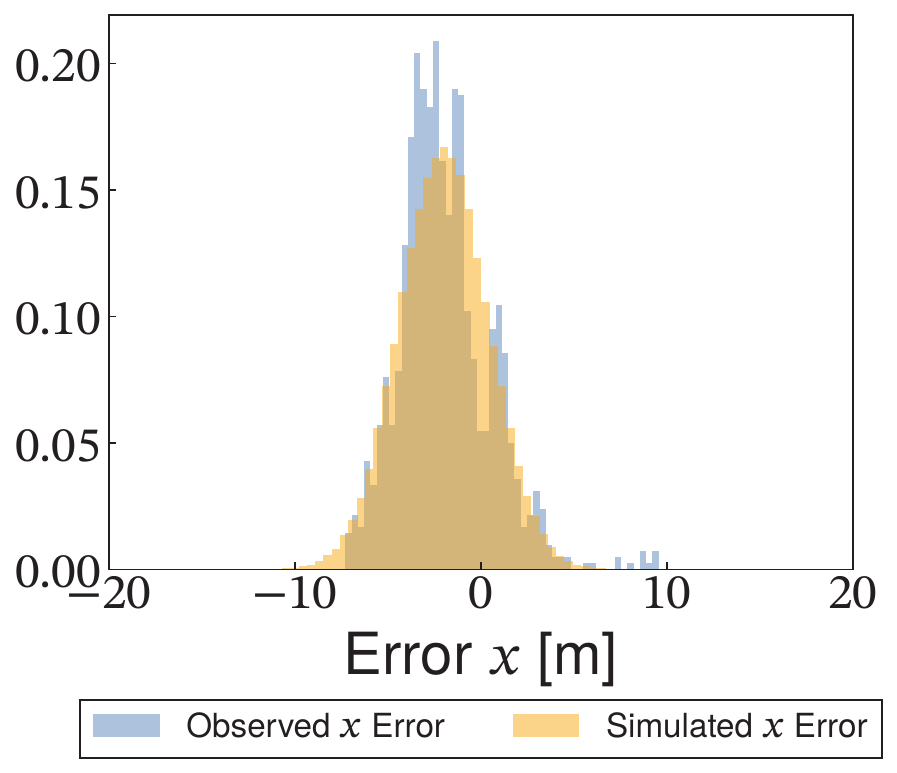}}
    \subfloat[]{\includegraphics[width=0.45\linewidth]{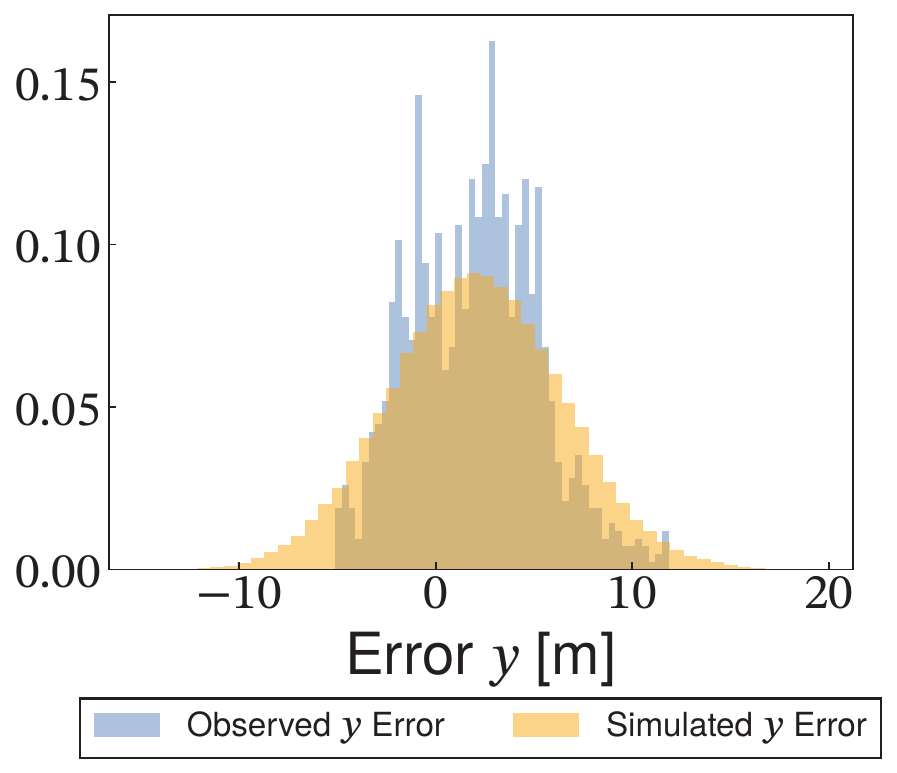}}
    \caption{Histograms of positioning error: (a) motorola edge 40 neo ($x$-axis), (b) motorola edge 40 neo ($y$-axis), (c) moto g24 ($x$-axis), (d) moto g24 ($y$-axis).}
    \label{fig:error_histogram}
\end{figure}

\begin{figure}[t!]
    \centering
    \subfloat[]{\includegraphics[width=0.45\linewidth]{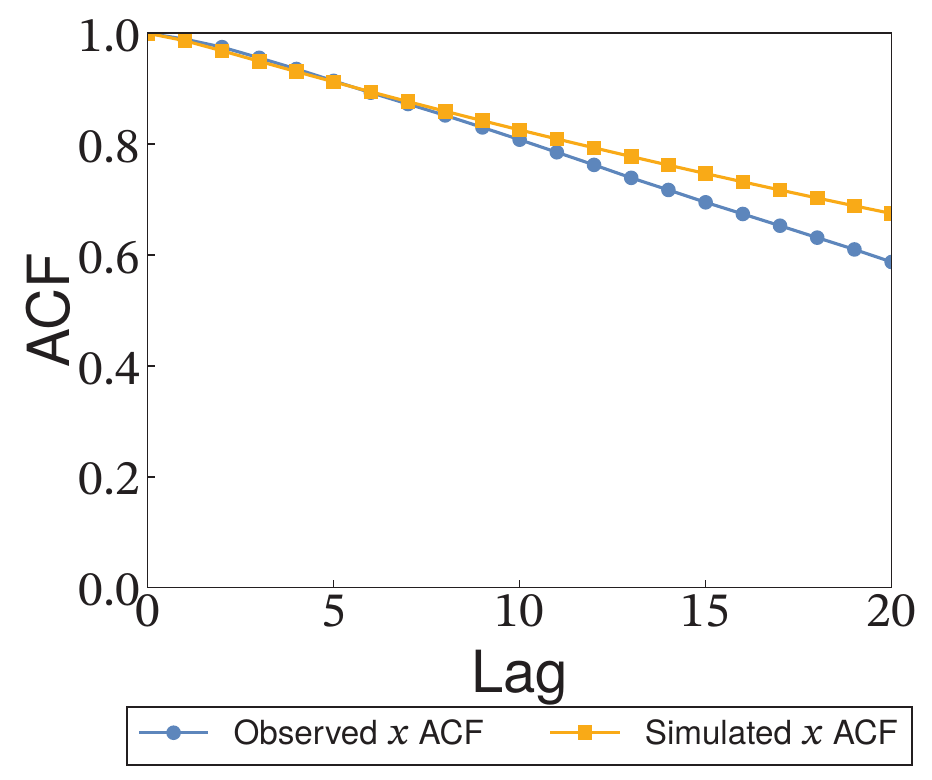}}
    \subfloat[]{\includegraphics[width=0.45\linewidth]{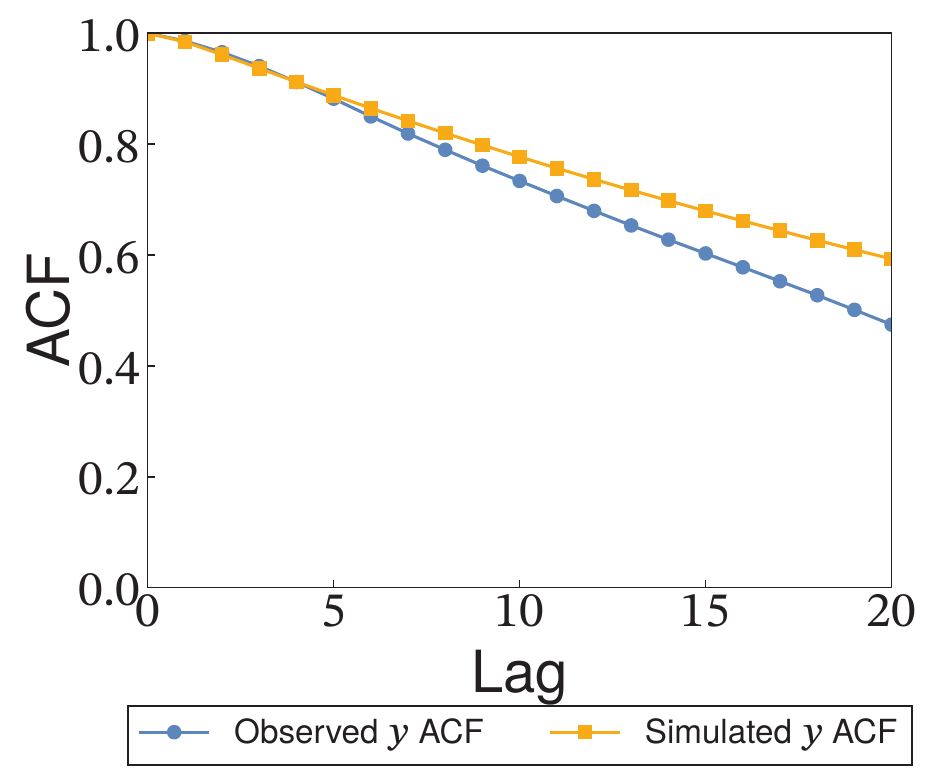}} \\
    \subfloat[]{\includegraphics[width=0.45\linewidth]{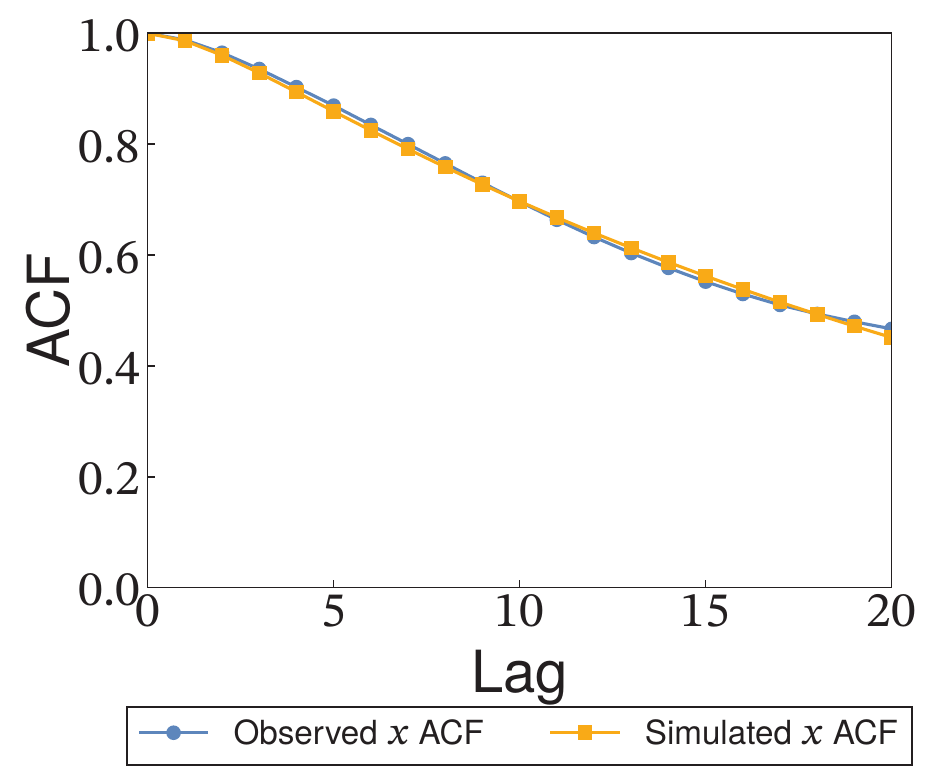}}
    \subfloat[]{\includegraphics[width=0.45\linewidth]{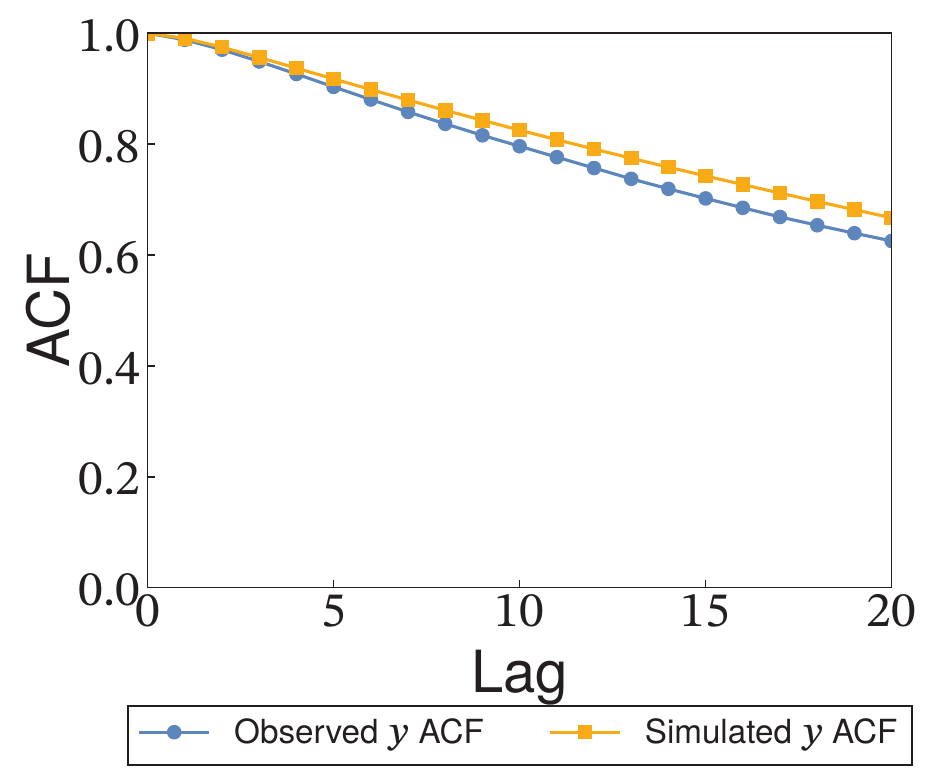}}
    \caption{ACF of positioning error (a) motorola edge 40 neo ($x$-axis), (b) motorola edge 40 neo ($y$-axis), (c) moto g24 ($x$-axis), (d) moto g24 ($y$-axis).}
    \label{fig:ACF}
\end{figure}

The GNSS measurements were collected on the campus of The University of Electro-Communications on June 6, 2025.
The reference trajectory was obtained by RTK positioning using a u-blox ZED-F9R, while the GNSS observation trajectories were collected using two Motorola smartphones, a motorola edge 40 neo and a moto g24.
The motorola edge 40 neo is equipped with a dual-frequency GNSS module, whereas the moto g24 uses a single-frequency GNSS module.
Positions were recorded at 5\,Hz by the ZED-F9R and at 1\,Hz by the smartphones, yielding 1249 observed positions from the motorola edge 40 neo and 1251 from the moto g24.
The measurement setup and observed trajectories are shown in Figs.\,\ref{fig:observations} and\,\ref{fig:observation_trajectory}, respectively.
The parameters of the error model in Eqs.\,\eqref{eq:gnss_error_decomposition} and\,\eqref{eq:gnss_ar2_model} identified for the two devices are summarized in Table\,\ref{tab:gnss_model_parameter}.
Here, $\mathbf{a}$ denotes the quasi-static offset obtained from each measured sequence, whereas the AR(2) coefficients and innovation standard deviations characterize the 1\,Hz time-varying residual.
\begin{table}[t!]
    \centering
    \caption{Parameters of 1\,Hz AR(2) models identified from measured GNSS positioning errors.}
    \label{tab:gnss_model_parameter}
    \begin{tabular}{ccc} \hline
        Parameter & motorola edge 40 neo & moto g24 \\ \hline
        $\mathbf{a}$ [m]
            & $[-6.37,\ 1.00]^\top$
            & $[-1.91,\ 1.99]^\top$ \\
        $\omega_{x,1}$
            & 1.183
            & 1.491 \\
        $\omega_{x,2}$
            & $-1.947\times 10^{-1}$
            & $-5.084\times 10^{-1}$ \\
        $\sigma_x$ [m]
            & $4.836\times 10^{-1}$
            & $3.130\times 10^{-1}$ \\
        $\omega_{y,1}$
            & 1.246
            & 1.405 \\
        $\omega_{y,2}$
            & $-2.621\times 10^{-1}$
            & $-4.144\times 10^{-1}$ \\
        $\sigma_y$ [m]
            & $5.258\times 10^{-1}$
            & $4.625\times 10^{-1}$ \\ \hline
    \end{tabular}
\end{table}
The histograms of the measured positioning errors and those generated from the identified AR(2) models are shown in Fig.\,\ref{fig:error_histogram}, and the corresponding autocorrelation functions (ACFs) are shown in
Fig.\,\ref{fig:ACF}.
Although differences between the measured and generated marginal distributions can be observed, the AR(2) models approximately reproduce both the strength and decay characteristics of the measured autocorrelation.
Thus, the AR(2) models are not intended to reproduce the complete distribution of the GNSS positioning errors; rather, they provide measurement-informed approximations of the temporal correlation considered in Sec.\,\ref{sec:real_gnss_evaluation}.
The measured sequences also show that strongly temporally correlated components remain after accounting for the quasi-static offset, motivating the model-mismatch evaluation in that section.
\subsection{Construction of the 20\,s AR(2) Model}
\label{app:ar2_20s_conversion}
The AR(2) models in Table\,\ref{tab:gnss_model_parameter} were identified from GNSS positioning-error sequences sampled at 1\,Hz, whereas the numerical evaluations in Sec.\,\ref{sec:numerical_evaluation} use a measurement interval of 20\,s.
Directly using the 1\,Hz AR(2) coefficients at the 20\,s sampling interval would not preserve the temporal correlation of the original process.
We therefore construct an approximate AR(2) model at 20\,s intervals by matching selected stationary autocovariances of the 1\,Hz model.

For each axis $q\in\{x,y\}$, let the stationary autocovariance of the 1\,Hz AR(2) process be
\begin{equation}
    \label{eq:ar2_autocovariance}
    \gamma_q(\ell) \triangleq \mathbb{E}\left[u_{q,k}u_{q,k-\ell}\right].
\end{equation}
The approximate AR(2) process at 20\,s intervals is written as
\begin{equation}
    \label{eq:projected_ar2}
    \bar{u}_{q,n} = \overline{\omega}_{q,1}\bar{u}_{q,n-1} + \overline{\omega}_{q,2}\bar{u}_{q,n-2} + \overline{\epsilon}_{q,n},  \qquad \bar{\epsilon}_{q,n} \sim \mathcal{N}(0,\bar{\sigma}_q^2).
\end{equation}
To preserve the stationary autocovariances at lags 0, 20, and 40\,s, the AR coefficients are obtained from the Yule--Walker
equations\,\cite{friedlanderModifiedYuleWalkerMethod1984},
\begin{equation}
    \label{eq:ar2_yule_walker_projection}
    \begin{bmatrix}
        \gamma_q(0)  & \gamma_q(20) \\
        \gamma_q(20) & \gamma_q(0)
    \end{bmatrix}
    \begin{bmatrix}
        \overline{\omega}_{q,1} \\
        \overline{\omega}_{q,2}
    \end{bmatrix}
    =
    \begin{bmatrix}
        \gamma_q(20) \\
        \gamma_q(40)
    \end{bmatrix}.
\end{equation}
The corresponding innovation variance is
\begin{equation}
    \label{eq:projected_ar2_innovation}
    \bar{\sigma}_q^2 = \gamma_q(0) - \overline{\omega}_{q,1}\gamma_q(20) - \overline{\omega}_{q,2}\gamma_q(40).
\end{equation}
This construction is a moment-matching approximation and does not assume that exact 20\,s subsampling of the original 1\,Hz AR(2) process remains an AR(2) process.
The resulting 20\,s models are used to generate the residual sequences $\overline{\mathbf{u}}_n^{(i)}$ in Eq.\,\eqref{eq:ar2_simulated_observed_position}.

\section{State-Space Formulation of the KF--RTS Baseline}
\label{app:kf_rts_details}
This appendix gives the complete state-space formulation of the KF--RTS baseline used in Sec.\,\ref{sec:real_gnss_evaluation}.
The baseline explicitly represents the temporally correlated AR(2) residual in a linear Gaussian state-space model, while using only the GNSS observation trajectory during the smoothing stage.
The resulting smoothed positions are subsequently used as the training locations for GPR.
For sensor $i$, the state vector at time $n$ is
\begin{multline}
    \label{eq:ar2_rts_state}
    \mathbf{s}_n^{(i)} = \left[ p_{x,n}^{(i)}, v_{x,n}^{(i)}, p_{y,n}^{(i)}, v_{y,n}^{(i)}, b_x^{(i)}, b_y^{(i)},\right. \\
    \left. u_{x,n}^{(i)}, u_{x,n-1}^{(i)}, u_{y,n}^{(i)}, u_{y,n-1}^{(i)} \right]^\top .
\end{multline}
Here, $\mathbf{p}_n^{(i)}=[p_{x,n}^{(i)},p_{y,n}^{(i)}]^\top$ and $\mathbf{v}_n^{(i)}$ denote the true position and velocity, respectively, $\mathbf{b}^{(i)}$ denotes the sensor-specific constant positioning bias, and $\mathbf{u}_n^{(i)}$ denotes the AR(2) residual. \par
For each axis $q\in\{x,y\}$, the true position and velocity follow the constant-velocity model
\begin{equation}
    \label{eq:rts_motion_model}
    \begin{bmatrix}
        p_{q,n}^{(i)}\\
        v_{q,n}^{(i)}
    \end{bmatrix}
    =
    \begin{bmatrix}
        1 & \Delta t\\
        0 & 1
    \end{bmatrix}
    \begin{bmatrix}
        p_{q,n-1}^{(i)}\\
        v_{q,n-1}^{(i)}
    \end{bmatrix}
    +
    \mathbf{w}_{q,n}^{(i)}.
\end{equation}
The process noise is modeled as
\begin{equation}
    \label{eq:rts_motion_process_noise}
    \mathbf{w}_{q,n}^{(i)} \sim \mathcal{N}\left( \mathbf{0}, q_{\mathrm{mot}}
        \begin{bmatrix}
            \Delta t^3/3 & \Delta t^2/2\\
            \Delta t^2/2 & \Delta t
        \end{bmatrix}
    \right),
\end{equation}
where $q_{\mathrm{mot}}=0.30\,\mathrm{m}^2/\mathrm{s}^3$ is used in this evaluation.
The constant positioning bias evolves according to
\begin{equation}
    \label{eq:rts_bias_transition}
    \mathbf{b}_n^{(i)} = \mathbf{b}_{n-1}^{(i)}.
\end{equation}
For the AR(2) residual, define $\mathbf{r}_{q,n}^{(i)} = [u_{q,n}^{(i)},u_{q,n-1}^{(i)}]^\top$.
Using the 20\,s AR(2) parameters obtained in Appendix\,\ref{app:gnss_model_details}, we write
\begin{equation}
    \label{eq:rts_ar2_transition}
    \mathbf{r}_{q,n}^{(i)} = \underbrace{
    \begin{bmatrix}
        \bar{\omega}_{q,1} & \bar{\omega}_{q,2}\\
        1 & 0
    \end{bmatrix}}_{\mathbf{A}_q} \mathbf{r}_{q,n-1}^{(i)} + \boldsymbol{\xi}_{q,n}^{(i)},
\end{equation}
where
\begin{equation}
    \label{eq:rts_ar2_process_noise}
    \boldsymbol{\xi}_{q,n}^{(i)} \sim \mathcal{N}\left( \mathbf{0}, \mathbf{Q}_q \right), \qquad \mathbf{Q}_q \triangleq
    \begin{bmatrix}
        \bar{\sigma}_q^2 & 0\\
        0 & 0
    \end{bmatrix}.
\end{equation}
When thinning increases the interval between consecutive retained measurements to $\Delta t=20K\,\mathrm{s}$, the state-transition matrix is $\mathbf{A}_q^K$, and the accumulated process covariance is
\begin{equation}
    \label{eq:rts_ar2_gap_covariance}
    \mathbf{Q}_{q,K} = \sum_{j=0}^{K-1} \mathbf{A}_q^j \mathbf{Q}_q \left(\mathbf{A}_q^j\right)^\top .
\end{equation}
The observation equation is
\begin{equation}
    \label{eq:rts_observation_model}
    \tilde{\mathbf{x}}_n^{(i)} =
    \begin{bmatrix}
        p_{x,n}^{(i)}+b_x^{(i)}+u_{x,n}^{(i)}\\
        p_{y,n}^{(i)}+b_y^{(i)}+u_{y,n}^{(i)}
    \end{bmatrix} + \boldsymbol{\nu}_n^{(i)}.
\end{equation}
Because the simulated observations in Eq.\,\eqref{eq:ar2_simulated_observed_position} contain no additional white positioning noise independent of the AR(2) residual, we set $\boldsymbol{\nu}_n^{(i)}=\mathbf{0}$. \par
For the initial position, the uniform distribution over the $300\,\mathrm{m}\times300\,\mathrm{m}$ simulation area is approximated by
a Gaussian distribution through moment matching,
\begin{equation}
    \label{eq:rts_initial_position}
    \mathbb{E}[\mathbf{p}_0] = [150,150]^\top, \qquad \operatorname{Cov}(\mathbf{p}_0) = 7500\mathbf{I}.
\end{equation}
We further use
\begin{equation}
    \label{eq:rts_initial_velocity_bias}
    \mathbb{E}[\mathbf{v}_0]=\mathbf{0}, \qquad \operatorname{Cov}(\mathbf{v}_0)=2\mathbf{I}, \qquad \mathbf{b}^{(i)} \sim \mathcal{N}(\mathbf{0},100\mathbf{I}),
\end{equation}
and initialize the AR(2) states from their stationary distributions. \par
A forward KF is first applied to the state-space model, followed by fixed-interval RTS smoothing over the entire observation sequence.
The smoothed positions $\hat{\mathbf{p}}_{n,\mathrm{RTS}}^{(i)}$ are then used as the training locations of the subsequent GPR.
In this comparison, the KF--RTS smoother is provided with the same 20\,s AR(2) coefficients and innovation variances as those used to generate the positioning errors.
However, the true positions, true constant biases, and realized AR(2) residuals in each trial are not provided to the baseline.
Only the GNSS observation trajectories are used for position smoothing, and the RSS observations are introduced only in the subsequent GPR stage. \par
The transition and observation likelihoods of this state-space model are invariant to the sensor-wise transformation $\mathbf{p}_n^{(i)}\mapsto \mathbf{p}_n^{(i)}+\mathbf{c}^{(i)}$ and $\mathbf{b}^{(i)}\mapsto \mathbf{b}^{(i)}-\mathbf{c}^{(i)}$ for any constant vector $\mathbf{c}^{(i)}\in\mathbb{R}^2$.
Therefore, apart from the absolute-position information introduced through the initial-position and bias priors, the GNSS trajectory likelihood alone cannot distinguish a constant translation of the trajectory from the corresponding sensor-specific positioning bias.
Even when the temporal AR(2) dynamics are correctly specified, temporal smoothing alone therefore provides no additional likelihood information for resolving this quasi-static translation ambiguity.
This is consistent with the trajectory-only identifiability discussion in
Sec.\,\ref{subsec:working_upper_bound}.

\bibliography{bibliography}

@book{goldsmithWirelessCommunications2005,
  title = {Wireless {{Communications}}},
  author = {Goldsmith, Andrea},
  year = 2005,
  month = aug,
  publisher = {Cambridge Univ. Press},
  isbn = {978-0-521-83716-3},
  langid = {english}
}

@article{gudmundsonCorrelationModelShadow1991,
  title = {Correlation Model for Shadow Fading in Mobile Radio Systems},
  author = {Gudmundson, M.},
  year = 1991,
  month = nov,
  journal = {Electron. Lett.},
  volume = {27},
  number = {23},
  pages = {2145--2146},
  publisher = {{The Institution of Engineering and Technology}},
  doi = {10.1049/el:19911328},
  urldate = {2025-09-08}
}

@article{xiongFaultTolerantGNSSSINS2020a,
  title = {Fault-{{Tolerant GNSS}}/{{SINS}}/{{DVL}}/{{CNS Integrated Navigation}} and {{Positioning Mechanism Based}} on {{Adaptive Information Sharing Factors}}},
  author = {Xiong, Hailiang and Bian, Ruochen and Li, Yujun and Du, Zhengfeng and Mai, Zhenzhen},
  year = 2020,
  month = sep,
  journal = {{IEEE} Syst. J.},
  volume = {14},
  number = {3},
  pages = {3744--3754},
  issn = {1937-9234},
  doi = {10.1109/JSYST.2020.2981366},
  urldate = {2025-12-30}
}

@article{rheeLevyWalkNatureHuman2011b,
  title = {On the {{Levy-Walk Nature}} of {{Human Mobility}}},
  author = {Rhee, Injong and Shin, Minsu and Hong, Seongik and Lee, Kyunghan and Kim, Seong Joon and Chong, Song},
  year = 2011,
  month = jun,
  journal = {{IEEE/ACM} Trans. Netw.},
  volume = {19},
  number = {3},
  pages = {630--643},
  issn = {1558-2566},
  doi = {10.1109/TNET.2011.2120618},
  urldate = {2026-04-12}
}

@article{tranNetworkDigitalTwin2025,
  title = {Network {{Digital Twin}} for {{6G}} and {{Beyond}}: {{An End-to-End View Across Multi-Domain Network Ecosystems}}},
  shorttitle = {Network {{Digital Twin}} for {{6G}} and {{Beyond}}},
  author = {Tran, Dinh-Hieu and Waheed, Nazar and Saputra, Yuris Mulya and Lin, Xingqin and Nguyen, Cong T. and Abdu, Tedros Salih and Vo, Van Nhan and Pham, Van-Quan and Alsenwi, Madyan and Adam, Abuzar Babikir Mohammad and Chatzinotas, Symeon and Lagunas, Eva and Tran, Hung and Dac, Tu Ho and Huynh, Nguyen Van},
  year = 2025,
  journal = {{IEEE} Open J. Commun. Soc.},
  volume = {6},
  pages = {6866--6911},
  issn = {2644-125X},
  doi = {10.1109/OJCOMS.2025.3599866},
  urldate = {2026-04-21}
}

@article{zengTutorialEnvironmentAwareCommunications2024b,
  title = {A {{Tutorial}} on {{Environment-Aware Communications}} via {{Channel Knowledge Map}} for {{6G}}},
  author = {Zeng, Yong and Chen, Junting and Xu, Jie and Wu, Di and Xu, Xiaoli and Jin, Shi and Gao, Xiqi and Gesbert, David and Cui, Shuguang and Zhang, Rui},
  year = 2024,
  journal = {{IEEE} Commun. Surveys Tuts.},
  volume = {26},
  number = {3},
  pages = {1478--1519},
  issn = {1553-877X},
  doi = {10.1109/COMST.2024.3364508},
  urldate = {2026-04-14}
}

@article{zhangWiFiBasedIndoorLocalization2024a,
  title = {Wi-{{Fi-Based Indoor Localization With Interval Random Analysis}} and {{Improved Particle Swarm Optimization}}},
  author = {Zhang, Xing and Sun, Wei and Zheng, Jin and Lin, Anping and Liu, Jian and Ge, Shuzhi Sam},
  year = 2024,
  month = oct,
  journal = {{IEEE} Trans. Mobile Comput.},
  volume = {23},
  number = {10},
  pages = {9120--9134},
  issn = {1558-0660},
  doi = {10.1109/TMC.2024.3359669},
  urldate = {2026-04-21}
}

@article{suarezrodriguezNetworkOptimisation5G2020,
  title = {Network {{Optimisation}} in {{5G Networks}}: {{A Radio Environment Map Approach}}},
  shorttitle = {Network {{Optimisation}} in {{5G Networks}}},
  author = {Suarez Rodriguez, Antonio Cristo and Haider, Noman and He, Ying and Dutkiewicz, Eryk},
  year = 2020,
  month = oct,
  journal = {{IEEE} Trans. Veh. Technol.},
  volume = {69},
  number = {10},
  pages = {12043--12057},
  issn = {1939-9359},
  doi = {10.1109/TVT.2020.3011147},
  urldate = {2026-04-21}
}

@article{biEngineeringRadioMaps2019b,
  title = {Engineering {{Radio Maps}} for {{Wireless Resource Management}}},
  author = {Bi, Suzhi and Lyu, Jiangbin and Ding, Zhi and Zhang, Rui},
  year = 2019,
  month = apr,
  journal = {{IEEE} Wireless Commun.},
  volume = {26},
  number = {2},
  pages = {133--141},
  issn = {1558-0687},
  doi = {10.1109/MWC.2019.1800146},
  urldate = {2026-04-21}
}

@article{chenGPRTGaussianProcess2025,
  title = {{{GPRT}}: {{A Gaussian Process Regression-Based Radio Map Construction Method}} for {{Rugged Terrain}}},
  shorttitle = {{{GPRT}}},
  author = {Chen, Guokai and Liu, Yongxiang and Zhang, Jianzhao and Zhang, Tao and Liu, Kai and Yang, Jun},
  year = 2025,
  month = jul,
  journal = {{IEEE} Internet Things J.},
  volume = {12},
  number = {13},
  pages = {23905--23920},
  issn = {2327-4662},
  doi = {10.1109/JIOT.2025.3554507},
  urldate = {2026-08-15}
}

@article{sunPropagationMapReconstruction2022,
  title = {Propagation {{Map Reconstruction}} via {{Interpolation Assisted Matrix Completion}}},
  author = {Sun, Hao and Chen, Junting},
  year = 2022,
  journal = {{IEEE} Trans. Signal Process.},
  volume = {70},
  pages = {6154--6169},
  issn = {1941-0476},
  doi = {10.1109/TSP.2022.3230332},
  urldate = {2026-08-15}
}

@article{teganyaDeepCompletionAutoencoders2022,
  title = {Deep {{Completion Autoencoders}} for {{Radio Map Estimation}}},
  author = {Teganya, Yves and Romero, Daniel},
  year = 2022,
  month = mar,
  journal = {{IEEE} Trans. Wireless Commun.},
  volume = {21},
  number = {3},
  pages = {1710--1724},
  issn = {1558-2248},
  doi = {10.1109/TWC.2021.3106154},
  urldate = {2026-08-15}
}

@article{zhuGNSSPositionIntegrity2018,
  title = {{{GNSS Position Integrity}} in {{Urban Environments}}: {{A Review}} of {{Literature}}},
  shorttitle = {{{GNSS Position Integrity}} in {{Urban Environments}}},
  author = {Zhu, Ni and Marais, Juliette and B{\'e}taille, David and Berbineau, Marion},
  year = 2018,
  month = sep,
  journal = {{IEEE} Trans. Intell. Transp. Syst.},
  volume = {19},
  number = {9},
  pages = {2762--2778},
  issn = {1558-0016},
  doi = {10.1109/TITS.2017.2766768},
  urldate = {2026-08-15}
}

@inproceedings{mchutchonGaussianProcessTraining2011a,
  title = {Gaussian {{Process Training}} with {{Input Noise}}},
  booktitle = {Adv. Neural Inf. Process. Syst.},
  author = {Mchutchon, Andrew and Rasmussen, Carl},
  year = 2011,
  volume = {24},
  publisher = {Curran Associates, Inc.},
  urldate = {2026-04-12}
}

@article{zhenRadioEnvironmentMap2022,
  title = {Radio {{Environment Map Construction Based}} on {{Gaussian Process With Positional Uncertainty}}},
  author = {Zhen, Pan and Zhang, Bangning and Xu, Yi-Qun and Chen, Zhibo and Wang, Heng and Guo, Daoxing},
  year = 2022,
  month = aug,
  journal = {{IEEE} Wireless Commun. Lett.},
  volume = {11},
  number = {8},
  pages = {1639--1643},
  issn = {2162-2345},
  doi = {10.1109/LWC.2022.3170147},
  urldate = {2026-08-16}
}

@article{damianouVariationalInferenceLatent2016,
  title = {Variational {{Inference}} for {{Latent Variables}} and {{Uncertain Inputs}} in {{Gaussian Processes}}},
  author = {Damianou, Andreas C. and Titsias, Michalis K. and Lawrence, Neil D.},
  year = 2016,
  journal = {J. Mach. Learn. Res.},
  volume = {17},
  number = {42},
  pages = {1--62},
  issn = {1533-7928},
  urldate = {2026-08-16}
}

@article{iyerEnhancingPositioningGNSS2024,
  title = {Enhancing {{Positioning}} in {{GNSS Denied Environments Based}} on an {{Extended Kalman Filter Using Past GNSS Measurements}} and {{IMU}}},
  author = {Iyer, Kaushik and Dey, Abhijit and Xu, Bing and Sharma, Nitin and Hsu, Li-Ta},
  year = 2024,
  month = jun,
  journal = {{IEEE} Trans. Veh. Technol.},
  volume = {73},
  number = {6},
  pages = {7908--7924},
  issn = {1939-9359},
  doi = {10.1109/TVT.2024.3360076},
  urldate = {2026-04-15}
}

@online{kingmaAdamMethodStochastic2017,
  title = {Adam: {{A Method}} for {{Stochastic Optimization}}},
  shorttitle = {Adam},
  author = {Kingma, Diederik P. and Ba, Jimmy},
  date = {2017-01-30},
  eprint = {1412.6980},
  eprinttype = {arXiv},
  eprintclass = {cs.LG},
  doi = {10.48550/arXiv.1412.6980},
  url = {http://arxiv.org/abs/1412.6980},
  urldate = {2026-08-16},
  pubstate = {prepublished}
}

@article{romeroRadioMapEstimation2022b,
  title = {Radio {{Map Estimation}}: {{A}} Data-Driven Approach to Spectrum Cartography},
  shorttitle = {Radio {{Map Estimation}}},
  author = {Romero, Daniel and Kim, Seung-Jun},
  year = 2022,
  month = nov,
  journal = {{IEEE} Signal Process. Mag.},
  volume = {39},
  number = {6},
  pages = {53--72},
  issn = {1558-0792},
  doi = {10.1109/MSP.2022.3200175},
  urldate = {2026-08-16}
}

@inproceedings{girardGaussianProcessPriors2002,
  title = {Gaussian {{Process Priors}} with {{Uncertain Inputs Application}} to {{Multiple-Step Ahead Time Series Forecasting}}},
  booktitle = {Adv. Neural Inf. Process. Syst.},
  author = {Girard, Agathe and Rasmussen, Carl and Candela, Joaquin Qui{\~n}onero and {Murray-Smith}, Roderick},
  year = 2002,
  volume = {15},
  publisher = {MIT Press},
  urldate = {2026-08-16}
}

@book{sarkkaBayesianFilteringSmoothing2023,
  title = {Bayesian {{Filtering}} and {{Smoothing}}},
  author = {S{\"a}rkk{\"a}, Simo and Svensson, Lennart},
  year = 2023,
  publisher = {Cambridge Univ. Press},
  googlebooks = {WLe9EAAAQBAJ},
  isbn = {978-1-108-92664-5},
  langid = {english}
}

@article{kalmanNewApproachLinear1960,
  title = {A {{New Approach}} to {{Linear Filtering}} and {{Prediction Problems}}},
  author = {Kalman, R. E.},
  year = 1960,
  month = mar,
  journal = {J. Basic Eng.},
  volume = {82},
  number = {1},
  pages = {35--45},
  issn = {0021-9223},
  doi = {10.1115/1.3662552},
  urldate = {2026-08-16}
}

@book{rasmussenGaussianProcessesMachine2008,
  title = {{Gaussian Processes for Machine Learning}},
  author = {Rasmussen, Carl Edward and Williams, Christopher K. I.},
  year = 2005,
  publisher = {MIT Press},
  address = {Cambridge, Mass.},
  isbn = {978-0-262-18253-9},
  langid = {英語}
}

@article{nealAnnealedImportanceSampling2001,
  title = {Annealed Importance Sampling},
  author = {Neal, Radford M.},
  year = 2001,
  month = apr,
  journal = {Stat. Comput.},
  volume = {11},
  number = {2},
  pages = {125--139},
  issn = {1573-1375},
  doi = {10.1023/A:1008923215028},
  urldate = {2026-08-16},
  langid = {english}
}

@article{delmoralSequentialMonteCarlo2006,
  title = {Sequential {{Monte Carlo Samplers}}},
  author = {Del Moral, Pierre and Doucet, Arnaud and Jasra, Ajay},
  year = 2006,
  month = jun,
  journal = {J. R. Stat. Soc. Ser. B Stat. Methodol.},
  volume = {68},
  number = {3},
  pages = {411--436},
  issn = {1369-7412},
  doi = {10.1111/j.1467-9868.2006.00553.x},
  urldate = {2026-08-16}
}

@article{chopinSequentialParticleFilter2002,
  title = {A {{Sequential Particle Filter Method}} for {{Static Models}}},
  author = {Chopin, Nicolas},
  year = 2002,
  journal = {Biometrika},
  volume = {89},
  number = {3},
  eprint = {4140600},
  eprinttype = {jstor},
  pages = {539--551},
  publisher = {[Oxford Univ. Press, Biometrika Trust]},
  issn = {0006-3444},
  urldate = {2026-08-16}
}

@article{friedlanderModifiedYuleWalkerMethod1984,
  title = {The {{Modified Yule-Walker Method}} of {{ARMA Spectral Estimation}}},
  author = {Friedlander, Benjamin and Porat, Boaz},
  year = 1984,
  month = mar,
  journal = {{IEEE} Trans. Aerosp. Electron. Syst.},
  volume = {AES-20},
  number = {2},
  pages = {158--173},
  issn = {1557-9603},
  doi = {10.1109/TAES.1984.310437},
  urldate = {2026-08-16}
}

@article{wangDigitalTwinChannel2025,
  title = {Digital {{Twin Channel}} for {{6G}}: {{Concepts}}, {{Architectures}} and {{Potential Applications}}},
  shorttitle = {Digital {{Twin Channel}} for {{6G}}},
  author = {Wang, Heng and Zhang, Jianhua and Nie, Gaofeng and Yu, Li and Yuan, Zhiqiang and Li, Tongjie and Wang, Jialin and Liu, Guangyi},
  year = 2025,
  month = mar,
  journal = {IEEE Commun. Mag.},
  volume = {63},
  number = {3},
  pages = {24--30},
  issn = {1558-1896},
  doi = {10.1109/MCOM.001.2400213},
  urldate = {2026-09-01}
}

@article{ranacherWhyGPSMakes2016,
  title = {Why {{GPS}} Makes Distances Bigger than They Are},
  author = {Ranacher, Peter and Brunauer, Richard and Trutschnig, Wolfgang and {Van der Spek}, Stefan and Reich, Siegfried},
  year = 2016,
  month = feb,
  journal = {Int. J. Geogr. Inf. Sci.},
  volume = {30},
  number = {2},
  pages = {316--333},
  publisher = {Taylor \& Francis},
  issn = {1365-8816},
  doi = {10.1080/13658816.2015.1086924},
  urldate = {2026-09-01},
  pmid = {27019610}
}

@inproceedings{kanzakiJointExPostLocation2025,
  title = {Joint {{Ex-Post Location Calibration}} and {{Radio Map Construction}} under {{Biased Positioning Errors}}},
  booktitle = {2025 {{IEEE Globecom Workshops}}},
  author = {Kanzaki, Koki and Sato, Koya},
  year = 2025,
  month = dec,
  pages = {2071--2076},
  issn = {2166-0077},
  doi = {10.1109/GCWkshps68340.2025.11591047},
  urldate = {2026-09-04}
}
\bibliographystyle{IEEEtran}

\begin{IEEEbiographynophoto}{Koki Kanzaki (Graduate Student Member, IEEE)}
  received the B.E. and M.E. degrees in engineering from The University of Electro-Communications in 2024 and 2026, respectively. He is currently pursuing a Ph.D. degree at the same university. He received first place in the ITU AI/ML in 5G Challenge 2025. His research interests include spatio-temporal statistics.
\end{IEEEbiographynophoto}

\begin{IEEEbiographynophoto}{Katsuya Suto (Senior Member, IEEE)}
  received the B.Sc. degree in computer engineering from Iwate University, Morioka, Japan, in 2011, and the M.Sc. and Ph.D. degrees in information science from Tohoku University, Sendai, Japan, in 2013 and 2016, respectively. He has worked as a Post-Doctoral Fellow for Research Abroad, Japan Society for the Promotion of Science, at the Broadband Communications Research Laboratory, University of Waterloo, Waterloo, ON, Canada, from 2016 to 2018. He is an Associate Professor with the Faculty of Information Science and Technology, Hokkaido University, Sapporo, Hokkaido. His research interests include semantic communication, radio propagation, deep learning, and graph representation. Dr. Suto is a member of IEICE. He received the Best Paper Award from the IEEE VTC2013-spring, IEEE/CIC ICCC2015, IEEE ICC2016, and IEEE Transactions on Computers in 2018.
\end{IEEEbiographynophoto}

\begin{IEEEbiographynophoto}{Koya Sato (Senior Member, IEEE)}
  received the B.E. degree in electrical engineering from Yamagata University, in 2013, and the M.E. and Ph.D. degrees from The University of Electro-Communications, in 2015 and 2018, respectively. From 2018 to 2021, he was an Assistant Professor with the Tokyo University of Science. He is currently an Assistant Professor with the Artificial Intelligence eXploration Research Center, The University of Electro-Communications. His current research interests include wireless communication, distributed machine learning, and spatial statistics. Since 2026, he has served as an Associate Editor of IEEE Open Journal of the Communications Society and IEICE Transactions on Communications.
\end{IEEEbiographynophoto}

\end{document}